\documentclass[useAMS,usenatbib]{mn2e}
\usepackage[dvips]{epsfig}
\include{psfig}
\newcommand{\be}{\begin{equation}}   
\newcommand{\ee}{\end{equation}}   
\newcommand{\bea}{\begin{eqnarray}}   
\newcommand{\eea}{\end{eqnarray}}

\title[The Multipole Vectors of WMAP...]{The Multipole Vectors of WMAP, and their frames 
and invariants}

\author[Kate Land and Jo\~{a}o Magueijo]{
Kate Land and Jo\~{a}o Magueijo 
\thanks{E-mail:kate.land@imperial.ac.uk, j.magueijo@ic.ac.uk}
\\\\
Theoretical Physics Group, Imperial College, Prince Consort Road, 
London SW7 2BZ, UK
}

\begin{document}

\date{\today}

\pagerange{\pageref{firstpage}--\pageref{lastpage}} \pubyear{2004}

\maketitle

\label{firstpage}

\begin{abstract}
We investigate the Statistical Isotropy and Gaussianity of the CMB fluctuations,
using a set of multipole vector functions capable of separating
these two issues. 
In general a multipole is broken into a frame and
$2\ell-3$ ordered invariants. The multipole frame is found to be suitably sensitive to 
galactic cuts. We then apply our method to real WMAP datasets; a coadded masked map, the 
Internal Linear Combinations map, and Wiener filtered and cleaned maps. 
Taken as a whole, multipoles in the range $\ell=2-10$ or $\ell=2-20$
show consistency with statistical isotropy, as proved by the Kolmogorov
test applied to the frame's Euler angles. This result in {\it not} inconsistent
with previous claims for a preferred direction in the sky for $\ell=2,...5$.
The multipole invariants also show overall consistency with Gaussianity
apart from a few anomalies of limited significance (98\%), listed at the
end of this paper.
\end{abstract}


\section{Introduction}\label{intro}
The recent WMAP data has enabled analysis of the CMB to much higher precision 
than ever before. The CMB acts like a curtain on the early universe and a 
back light on later times. It is also the largest thing we can observe, and 
therefore is a strong probe of features on cosmological scales.
Inflationary theories predict the CMB fluctuations to be Gaussian and 
Statistically Isotropic (SI); the WMAP
data should be the best way to test  these predictions. Unfortunately, 
there have been reports of quite severe anomalies in this respect~\citep{coles,copi1,
cruz,dondon,erik1,erik2,erik3,hansen1,hansen2,virgo,
us1,us3,larson,oliv,vielva,park,copi2}.

Specifically, significant hemisphere asymmetries have been detected\citep{erik1,hansen2,us1}, as have 
multipole alignments\citep{teg,oliv,copi1,copi2,us3}, implying the onset of a preferred direction.
Such statistical anisotropy would be in conflict with inflationary 
theories, in the sense that any inflation model which left the universe just large 
enough to observe any anisotropy induced by, for example, a non-trivial 
topology~\citep{riaz,teg,dodec,sperg}, would require extreme fine tuning. 
This aside, it is interesting 
to consider possible origins of anisotropy in the universe: primordial 
magnetic fields, strings or walls~\citep{arj}, or an intrinsically inhomogeneous
Universe~\citep{moffatinh}.

In this paper we build on the work of~\cite{teg,copi1,us2}, re-examining the issue
of multipole alignment in the light of a formalism based on the Maxwell multipole
vectors. These are an elegant way of extracting the extra degrees of freedom
beside the power spectrum contained in a given multipole. As explained in~\cite{us2},
it is important to convert these vectors into $2\ell-3$ invariants (a generalisation
of the concept of eigenvalue) and an orthonormal triad of vectors (a generalisation
of the concept of eigenvectors of a hermitian matrix). Such an operation neatly
separates the issues of non-Gaussianity and of violations of statistical isotropy.
We proposed such a scheme in~\cite{us2}, and a further study appeared in~\cite{dennis2},
with analytical distributions provided. 

The plan of our paper is as follows. In Section~\ref{sig} we distinguish
SI and non-Gaussianity in real and harmonic (Fourier) space. Then in Section~\ref{mult}
we review previous work done with multipole vectors, their invariants and frames, and
present our new proposal. In Section~\ref{mask} we demonstrate the effectiveness of the method 
with simulations, and we apply these techniques to a variety of renditions of the
WMAP data in Section~\ref{results}. In the face of some concrete problems raised by real 
life data we conclude with
an appraisal of our method, and how it stands in light of recent
claims for anomalies in Section~\ref{end}. 
This paper is to be seen as a technical companion paper to~\cite{us3}
where the focus is more heavily placed on the analysis of the reported anomalies.

\section{Statistical Isotropy and Gaussianity}\label{sig}


First we outline the definitions and implications of Statistical 
Isotropy and Gaussianity in terms of the terminology of Real 
and Fourier space.

\subsection{Real space}\label{real}
We consider the the temperature of the CMB photons received from 
different directions, $\hat{n}$, in terms of the fluctuation
$\Delta T(\hat{n})=T(\hat{n})-\bar{T}$. 
Due to the random nature of the quantum fluctuations that seeded the 
initial perturbations (according to current standard cosmological models),
our theories of this process can only determine 
the \emph{statistical} behaviour of 
 $\Delta T(\hat{n})$. A theory can only make definite 
predictions for \emph{ensemble averages}: averages taken over an infinity of 
universes.

One way to characterise all the statistical information of the $\Delta T$ field 
is through the n-point correlation functions:
\be
\langle \Delta T(\hat{n_1} ) \Delta T(\hat{n_2} )... \Delta T(\hat{n_n}) \rangle
\ee
Note that without SI, these functions depend on $\hat{n_1}, \hat{n_2},...,\hat{n_n}$.

A function, f({\bf r}), is statistically isotropic if its 
statistical properties are invariant under translations ${\bf r} \rightarrow 
{\bf r}+\delta{\bf r}$, and there is no 
preferred direction \citep{BiPS}. On the surface of a sphere, this is 
equivalent to the statistical properties being rotationally invariant. 
Statistical Isotropy of the temperature fluctuations therefore demands that the 
n-point correlation functions are invariant under rotations. They now only depend 
on the magnitude of the separation between the vectors, $|\hat{n_i}-\hat{n_j}|$.
For example, 
the 2-point function only depends on the distance between the 
2 vectors, or equivalently the dot product:
\be
\langle \Delta T(\hat{n_1}) \Delta T(\hat{n_2}) \rangle = 
C(\hat{n_1}.\hat{n_2})
\ee

As mentioned, our models of the early universe can make predictions 
for the statistical 
properties of the field $\Delta T(\hat{n})$. The simplest models of inflation 
predict the $\Delta T(\hat{n})$ to be a Gaussian random field. 
For a Gaussian field the n-point correlation functions are zero for 
odd n, and can be expressed in terms of the 2-point correlation function 
for even n. Therefore, the two-point correlation function contains all 
the statistical information for a Gaussian field.

The combination of SI and Gaussianity means the two point correlation function 
contains all the information AND it is rotationally invariant.

\subsection{Fourier Space}\label{SH}
The usual approach to analysing CMB data, is to expand the function 
$\Delta T(\hat{n})$ in terms 
of the spherical harmonics, and therefore work with the spherical 
harmonic coefficients; $a_{\ell m}$.
\be
\Delta T(\hat{n}) = \sum_{\ell,m} a_{\ell m} Y_{\ell m}(\hat{n})
\ee

We can translate properties of the function $\Delta T(\hat{n})$ into 
properties of the $a_{\ell m}$.
As mentioned in Section~\ref{real}, SI of $\Delta T(\hat{n})$ 
requires the 2-point correlation function of $\Delta T(\hat{n})$ to be 
rotationally invariant. The consequence for the $a_{\ell m}$ is that the 
2-point correlation function be of the form \citep{hu}:
\be
\langle a_{\ell_1 m_1} a_{\ell_2 m_2} \rangle = C_{\ell_1} \delta_{{\ell_1}{\ell_2}} \delta_{{m_1}{m_2}}
\label{power}
\ee
We call $C_\ell$ the power spectrum.

Similarly, requiring the 3-point correlation function in real space to be SI demands that 
the 3-point function of the $a_{\ell m}$s be of the form \citep{hu}:
\be
\langle a_{\ell_1 m_1} a_{\ell_2 m_2} a_{\ell_3 m_3} \rangle = 
\left
( \begin{array}{ccc} \ell_1 & \ell_2 & \ell_3 \\ m_1 & m_2 & m_3
\end{array} 
\right ) B_{\ell_1 \ell_2 \ell_3}
\ee
We call $B_{\ell_1 \ell_2 \ell_3}$ the Bi-spectrum.

Gaussianity of the temperature fluctuations $\Delta T(\hat{n})$ 
demands that the coefficients, $a_{\ell m}$, are 
gaussian random variables. Therefore, all the information about the 
field is contained in the 2-point correlation function 
$\langle a_{\ell_1 m_1} a_{\ell_2 m_2} \rangle$. All higher n-point 
correlations functions are to be obtained from the 2-point
function by means of Wick's theorem.

Therefore, in Fourier space terminology, the combination of SI and Gaussianity 
means the power spectrum $C_\ell$ 
contains all the information of the field.

\section{Multipoles and the Multipole Vectors}\label{mult}
A multipole, $T_\ell$, represents all the fluctuations in $\Delta T$ 
of a certain scale, $\ell \sim 1/\theta$. That is:
\bea
\Delta T(\hat{n}) &=& \sum_\ell{ T_\ell(\hat{n})}\\
T_\ell(\hat{n}) &=& \sum_m{a_{\ell m} Y_{\ell m}(\hat{n})}
\eea
The reality of $\Delta T$ means $a_{\ell -m}$ is determined by $a_{\ell m}$ 
and {\rm Im}$(a_{\ell 0})=0$. There are therefore $2\ell+1$ degrees of 
freedom in multipole $T_\ell$.

As discussed, SI demands that the n-point correlation functions be 
rotationally invariant, and in Fourier space this has consequences for the 
form the n-point functions 
$\langle a_{\ell_1 m_1} .... a_{\ell_n m_n} \rangle$ 
can take. 
In Equation~\ref{power} we can see that SI demands that the $a_{\ell m}$s depend 
only on $\ell$ (no m-preference) and are independent for different $\ell$ 
(no inter-$\ell$ correlations).
If we think about it in terms of the p.d.f of each $a_{\ell m}$, the 
assumption of Gaussianity means that the shape of these p.d.f are Gaussian, and the 
assumption of SI means that their scaling are the same for the $a_{\ell m}$s of 
the same multipole, and independent for different multipoles.
 In fact, one can see that m-preference 
would indicate a preferred frame because the $a_{\ell m}$s are not frame 
independent, they depend on the frame of the Fourier Transform. 
Therefore, any m-preference would be related to a particular frame.

\subsection{Separation of degrees of freedom for the quadrupole}\label{sep}
A multipole is an irreducible representation of the SO(3) group.
 Therefore, from its $2\ell+1$ degrees of freedom we should be able to 
extract 3 rotational degrees of freedom, {\it i.e.}, an orthonormal 
frame. SI requires no inter-$\ell$ correlations, and therefore the orthonormal 
frames of each multipole are independent. Equivalently, SI requires these frames to be
 uniformly distributed. 
The remaining $2\ell-2$ degrees of freedom (d.o.f) 
will be invariants. Of these, one can be the 
overall power in the multipole $C_{\ell}$ (a Gaussian d.o.f). The 
remaining $2\ell-3$ invariants would probe m-preference and the Gaussianity of 
that multipole (these invariants should be independent for different multipoles).

Therefore, if one separates the 3 rotational d.o.f from the invariants 
one can test SI in the form of no inter-$\ell$ correlations 
independently from the remaining issues of m-preference and Gaussianity. 

For the example of the quadrupole there is a simple way of separating 
the d.o.f \citep{conf}:
\be\label{quad1}
\delta T_2=Q_{ij}x^i x^j
\ee
where $Q_{ij}$ is a symmetric traceless matrix. From this
matrix one may extract three eigenvectors
and two independent combinations of invariant eigenvalues $\lambda_i$ 
(they sum to 0). The eigenvalues d.o.fs are essentially the power
spectrum $C_2$ (related to the sum of the squares of $\lambda_i$)
and the bispectrum $B_{2,2,2}$ (related to the determinant of the matrix, or
the product of $\lambda_i$). The eigenvectors provide us with the 
orthonormal frame. 

\subsection{A general multipole}\label{multvec}

In \cite{us2} a new method was proposed that generalises this construction
for any multipole.  It built on the work of \cite{copi1} who first introduced 
the Multipole Vector formalism in CMB analysis. They showed 
that a multipole can be represented in terms of $\ell$ unit vectors and 
an overall magnitude. The method has received much attention, from 
mathematicians securing and exploring features of the formalism 
\citep{dennis1,dennis2,katz,lrey,weeks}, and from astrophysicists 
applying it to the WMAP data \citep{copi1,copi2}. 
In fact the formalism was first discovered by Maxwell \citep{maxwell}. 
He showed that for a real function $f(x,y,z)$ which is an eigenfunction 
of the Laplacian on the unit sphere with eigenvalue $-\ell(\ell+1)$ 
({\it i.e.}, a spherical harmonic $Y_{\ell m}$), there exists 
$\ell$ unit vectors ${\bf v}_1,...,{\bf v}_\ell$ such that:
\be
f(x,y,z)=\nabla_{{\bf v}_1}...\nabla_{{\bf v}_\ell}\frac{1}{r}
\ee
where $\nabla_{{\bf v}_1}= {\bf v}_1.\nabla$ is the directional 
derivative operator, and $r=\sqrt{x^2+y^2+z^2}$. A more useful form 
of the representation is given by \citep{dennis1}:
\be
f({\bf r})=A({\bf r}.{\bf v}_1)...({\bf r}.{\bf v}_\ell) + r^2R
\label{easyvec}
\ee
where is $A$ is a constant, and R is a complicated term fully defined by the 
components of the ${\bf v}_i$.

With the Multipole Vector method, the $2\ell+1$ d.o.f are 
separated into $\ell$ unit vectors, $v_i$, and one 
invariant, A. In \cite{us2} we proposed a further step of 
selecting 2 of the vectors as ``anchor vectors'' (we discuss this 
step further in Section~\ref{anchors}). You then 
use these anchor vectors ($L_1, L_2$) to define an orthonormal frame:
\bea
x=\widehat{(L_1-L_2)}\\
y=\widehat{(L_1+L_2)}\\
z=\widehat{(L_1\times L_2)}
\eea
and $2\ell-3$ invariants:
\bea
X_{12}=L_1.L_2\\
X_{1j}=L_1.L_j\\
X_{2j}=L_2.L_j
\eea
for $j \geq 3$.
We also showed how the power spectrum and bispectrum can be written in 
terms of the $X_{ij}$ and A.

Note that the Multipole Vectors are ``headless'' vectors: they 
do not have a defined direction. This can be clearly seen in Equation~\ref{easyvec} 
where any change in the sign of a vector can be absorbed into the scalar $A$. 
Therefore, we have to be careful in the definition of our orthonormal frame. 
We impose possible directions on the ``anchor vectors'' by insisting that their directions 
satisfy $X_{12}\equiv L_1.L_2 > 0$. Therefore, these vectors can still point 
in either direction but their directions are not independent, and there 
are only 2 possible combinations. The $z$-axis is now uniquely defined, and 
the $x$, and $y$-axes are up to a rotation by $\pi$ around the $z$-axis. We simply 
reduce the allowed range for the Euler angles of the axes.

We now have an orthonormal frame defined by our anchor vectors; we call this 
the Multipole Frame. The remaining vectors'
information is contained in their angles. 
The minimum number of dot products needed to 
contain all the information about the vector positions is the $2\ell-3$ 
invariants $X_{ij}$. So we see that information about m-preference and 
Gaussianity are contained in the angles between vectors, i.e. their 
relative spread. 

Gaussianity and SI determine the
probability distribution of these angles (see~\cite{dennis2} for an analytical 
expression). 
Departures from the expected spread would indicate a departure from SI 
and/or Gaussianity. It is not possible to probe the 
issues of m-preference and non-Gaussianity separately, however in some 
extreme cases of m-preference the vectors form clear structures. For pure m-preference 
there is a frame where $(\ell-m)$ of the vectors will line along the z-axis of this 
frame, and the remaining m vectors will form a 2m regular polygon in the x-y plane. 
We refer to this as a ``Handle and Disc'' structure, with the aligning vectors 
making the ``Handle'', and the planar vectors forming the ``Disc''.

Note that this method involves {\emph all} the information contained in the 
multipoles, as opposed to methods that explore just one statistic 
{\it e.g.}, $B_{\ell \ell \ell}$. This obviously has many advantages as it 
allows for a much more thorough investigation.

\begin{figure*}
\centering
\hfill
\begin{minipage}{55mm}
\begin{center}
\psfig{file=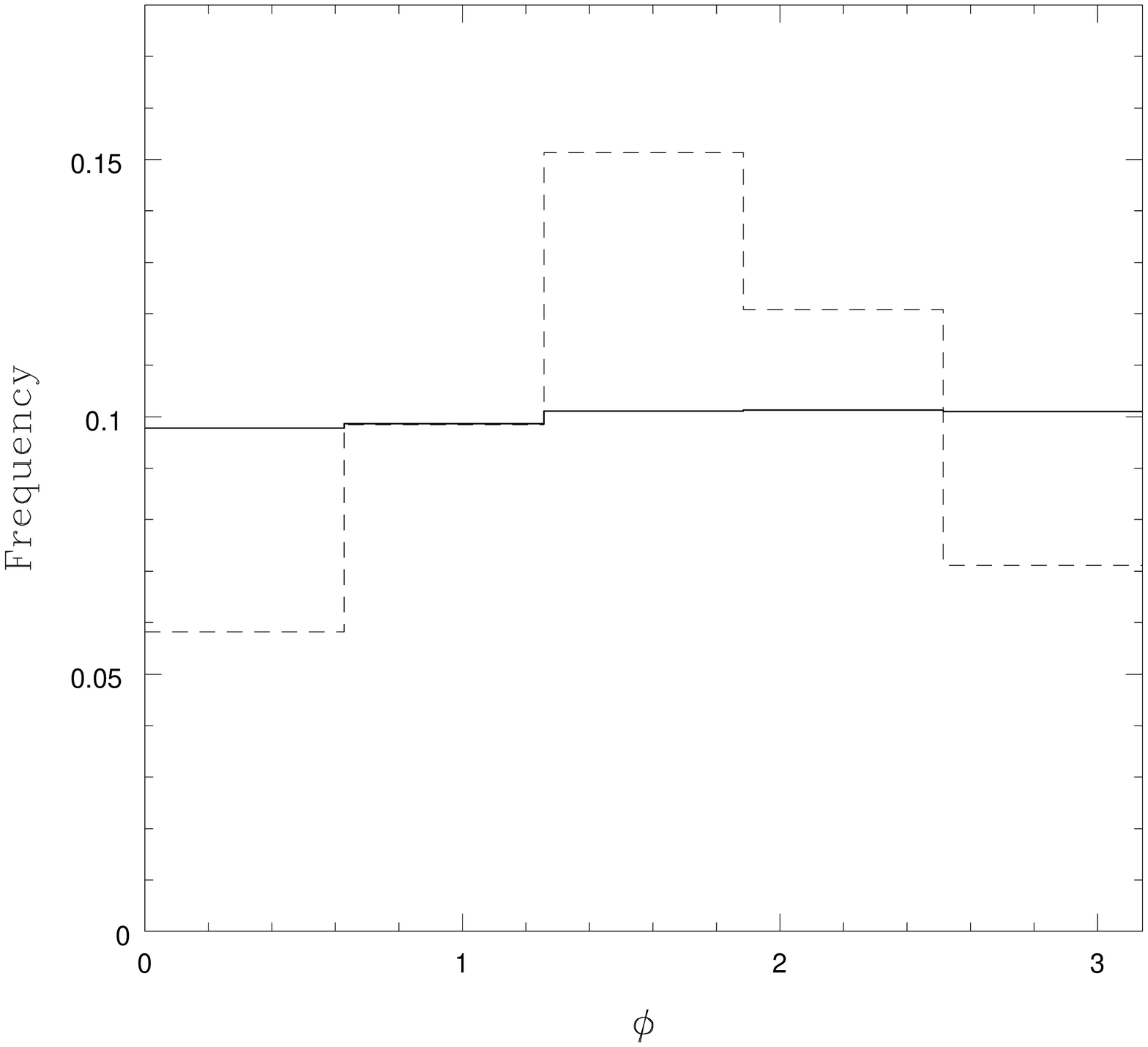,width=5.5cm}
\end{center}
\end{minipage}%
\hfill
\begin{minipage}{55mm}
\begin{center}
\psfig{file=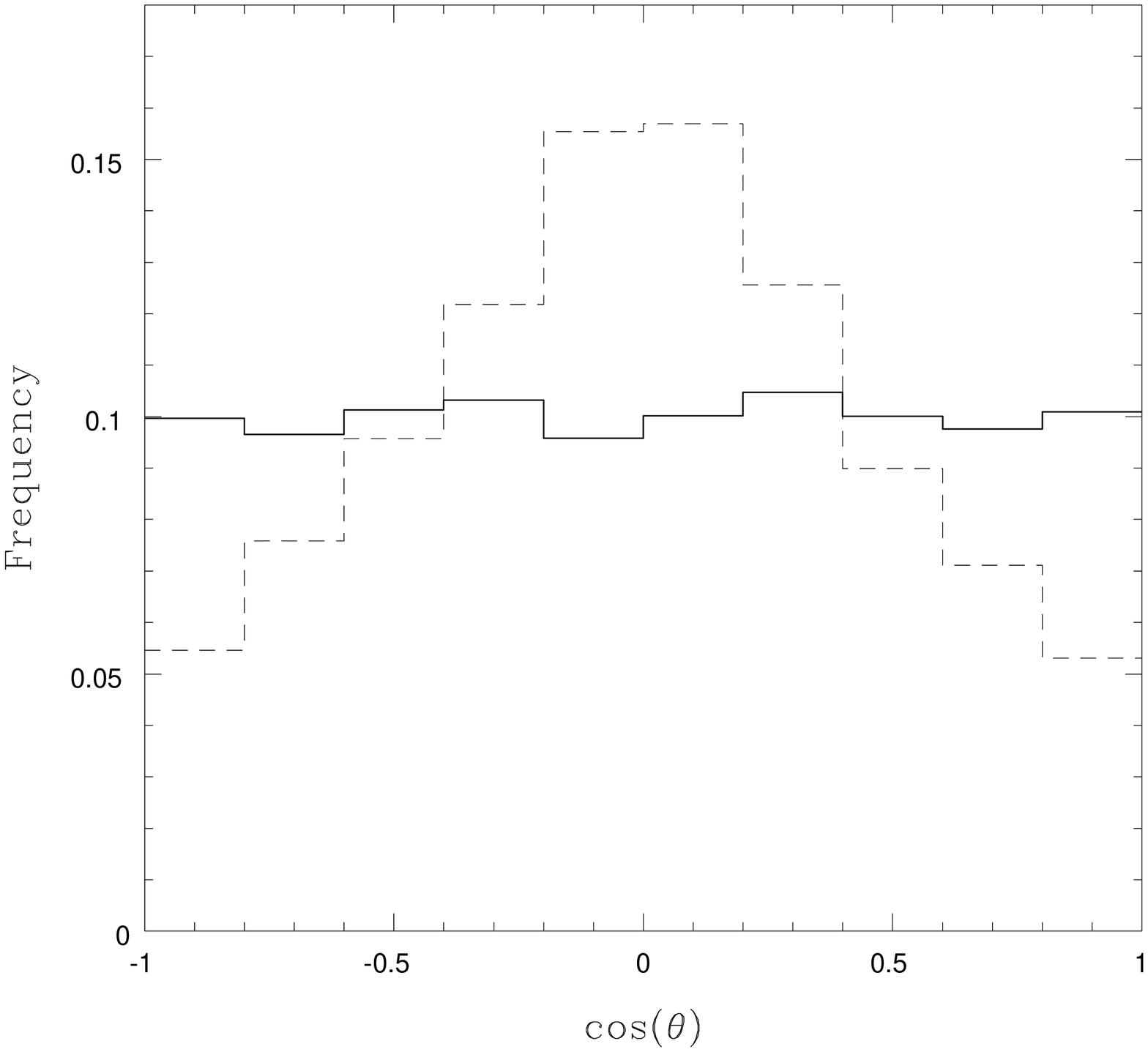,width=5.5cm}
\end{center}
\end{minipage}%
\hfill
\begin{minipage}{55mm}
\begin{center}
\psfig{file=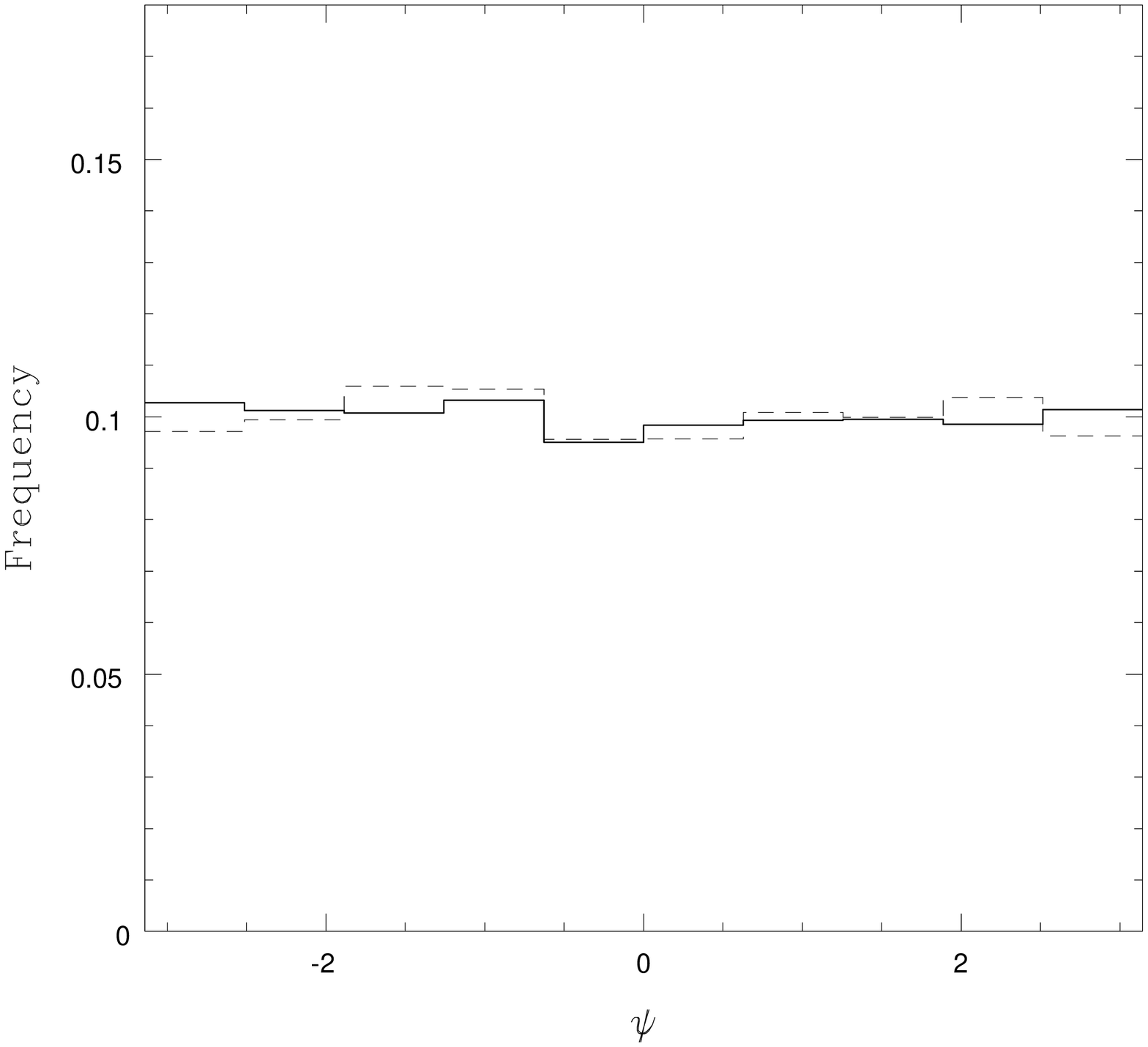,width=5.5cm}
\end{center}
\end{minipage}
\hfill
\caption{The Euler angles found for the Multipole frames of 1000 simulations, for multipoles $\ell=2-20$. 
The solid line is for the whole sky, and the dashed line is with the Kp0 masked applied.}
\label{MASKeuler}
\end{figure*}

\subsection{Anchors}\label{anchors}
We now explain further how to select the anchor vectors essential for our separation
of SI and non-Gaussian issues. 
We need an internal way of selecting these, {\it i.e.}, a method that does not depend on any 
outside coordinate frame. However, we want the process to be repeatable so 
selecting them at random (as we've done before) is not satisfactory. For instance
we may want to compare results for different frequency channels, so that the
algorithm must be unambiguous as well as invertible.

One such way would be to define the vectors as 
the two most aligned, or most orthogonal. This would select a pair, 
but would not provide 
them or the remaining vectors with an order. This is not a 
problem in defining the frame. However, with no ordering of the ${X_{1j},X_{2j}}$ 
we cannot assess the agreement of individual values with the simulations; we 
can only study them as a set. Ideally we want to use every piece of 
information, and therefore we look for a way to order the vectors.

We achieve this in the following way. From 
the set of $\ell$ unordered vectors $\{v_i\}$,  for every 
vector we sum the square of its dot products with all other vectors:
\be
K_i = \sum_j(X_{i,j})^2
\ee
where $X_{ij}=v_i.v_j$. Every vector now has a corresponding value of $K_i$, 
and we can order them. For example let $L_1$ be the vector 
with the maximum $K_i$ value, and $L_2$ equal the vector with the second 
highest $K_i$ value, etc... so $L_\ell$ is the vector with the lowest $K_i$ 
value. We can now calculate the $2\ell-3$ $X_{ij}$ values, and use the 
vectors $L_1$ and $L_2$ to define the frame.

\section{A simple application}\label{mask}

A sky cut ({\it e.g.}, imposed by the Kp0 mask) induces Anisotropy onto the sky.
 To investigate the effectiveness of our method we look at how 
well this Anisotropy is detected. We simulate 1000 Gaussian skies
with noise and beam, and we 
look at the orthonormal frames before and after we apply the Kp0 mask. 
We find the multipole frame $\{ {\bf x},{\bf y},{\bf z} \}$, as defined above, 
for multipoles $\ell=2-20$, and record the corresponding 
Euler angles $(\phi, \theta, \psi)$
\footnote{We are using the ``x-convention'', 
where the Euler angles $(\phi, \theta, \psi)$ represent first a rotation by 
$\phi$ around the $z$-axis, then by $\theta$ around the $x$-axis, 
then by $\psi$ around the $z$-axis (again). Further explanation 
can be found at 
{\it MathWorld}: http://mathworld.wolfram.com}. 
As mentioned in Section~\ref{multvec}, due to the Multipole Vectors having 
no defined direction, the definition of the orthonormal axes is 
defined up to $({\bf x},{\bf y})\rightarrow ({\bf -x},{\bf -y})$. 
Therefore there is an equivalence between orthonormal axes 
$\{ {\bf x},{\bf y},{\bf z} \} \sim \{ -{\bf x},-{\bf y},{\bf z} \}$, which 
translates into an equivalence between Euler angles 
$(\phi, \theta, \psi)\sim (\phi+\pi, \theta, \psi)$. Thus the 
allowed ranges for the Euler angles of our Multipole Frames are 
$\phi \in [0,\pi], \theta \in [0,\pi], \psi \in [-\pi, \pi]$.

In Figure~\ref{MASKeuler} we 
plot histograms of the distributions found for the Euler angles. We see that 
without any sky-cut the distributions are uniform as expected. However, 
we also clearly see how the sky-cut then does induce anisotropy. 
The $\phi$ and $cos(\theta)$ distributions become non-uniform:
$\theta$ peaks around $\pi/2$ ($z$-axis in galactic plane), 
and $\phi$ also around
$\pi/2$ ($x$-axis aligns with galactic poles). 
This is unlike other measures, such as the bispectrum, which  doesn't get largely 
effected by sky cuts.

In fact, one can predict this result. In terms of the Fourier expansion 
in spherical harmonics, the effect of the mask is to remove power 
from the $m\sim\pm\ell$ in galactic coordinates. Therefore we are 
equivalently inducing m-preference towards the lower $m$'s. As discussed in 
Section~\ref{multvec}, pure $m$-preference leads to a particular ``Handle and Disc'' 
configuration for the vectors, with $\ell-m$ vectors in the handle and $m$ 
in the disc. Therefore, for this induced preference towards the lower $m$'s 
we would expect a handle and disc structure (if only a subtle one) with most 
of the vectors in the handle. Our method of 
selecting the anchor vectors will then favour those in this handle as they will 
maximise $K_i$ through their small inter-angles with other handle members. 
If the anchor vectors belong to the handle, then the $x$-axis will 
align with this handle and therefore point to the galactic poles, 
the $y$-axis will not be well defined, and the 
$z$-axis will be perpendicular to the handle and therefore lying in the 
galactic plane.

\begin{figure*}
\vspace{5mm}
\hfill
\begin{minipage}{55mm}
\centerline{\psfig{file=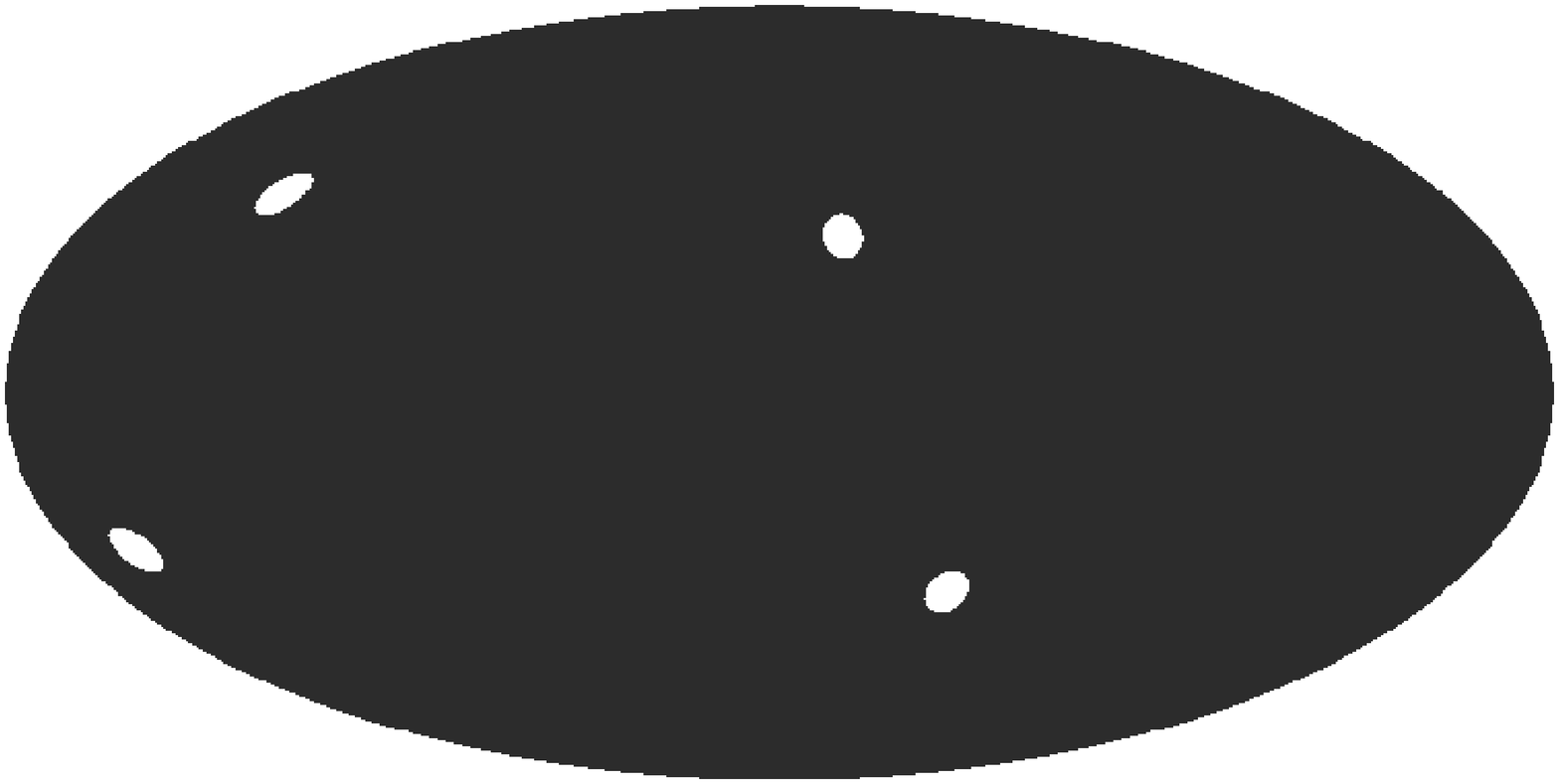,angle=90,width=4.5cm}}
\center{$\ell=2$}
\end{minipage}%
\hfill
\begin{minipage}{55mm}
\centerline{\psfig{file=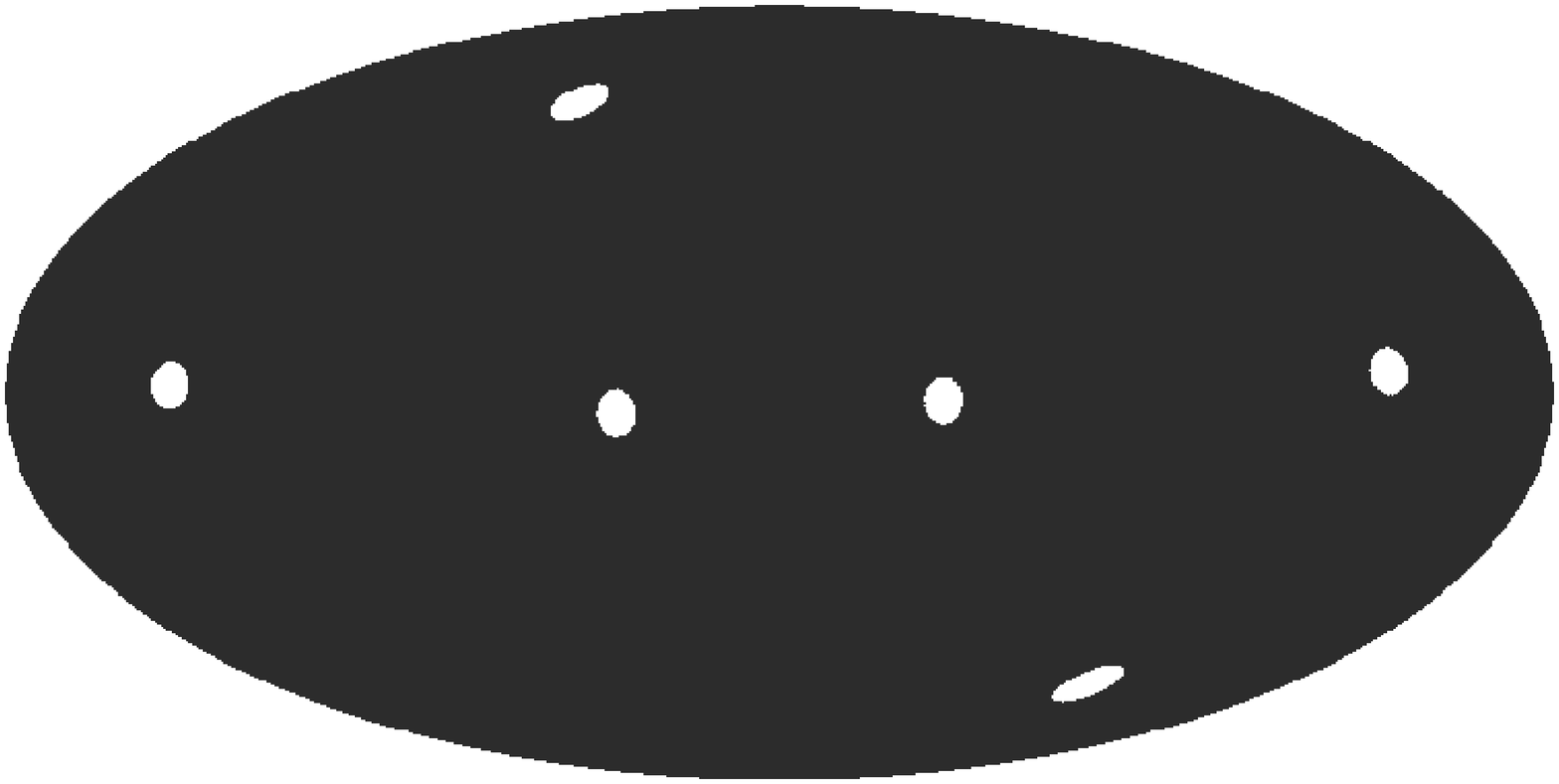,angle=90,width=4.5cm}}
\center{$\ell=3$}
\end{minipage}%
\hfill
\begin{minipage}{55mm}
\centerline{\psfig{file=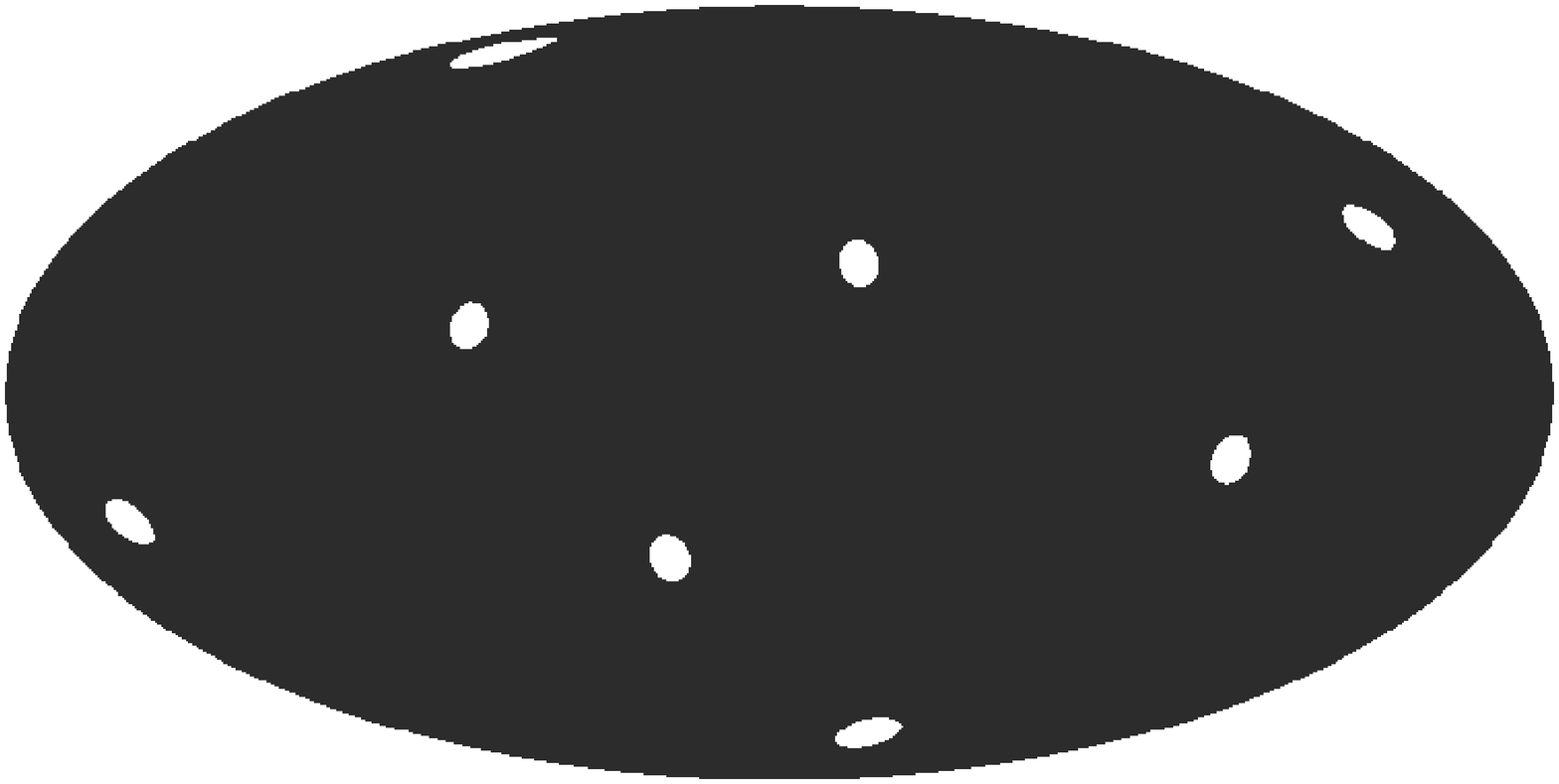,angle=90,width=4.5cm}}
\center{$\ell=4$}
\end{minipage}%
\hfill
\vspace{6mm}
\hfill
\begin{minipage}{55mm}
\centerline{\psfig{file=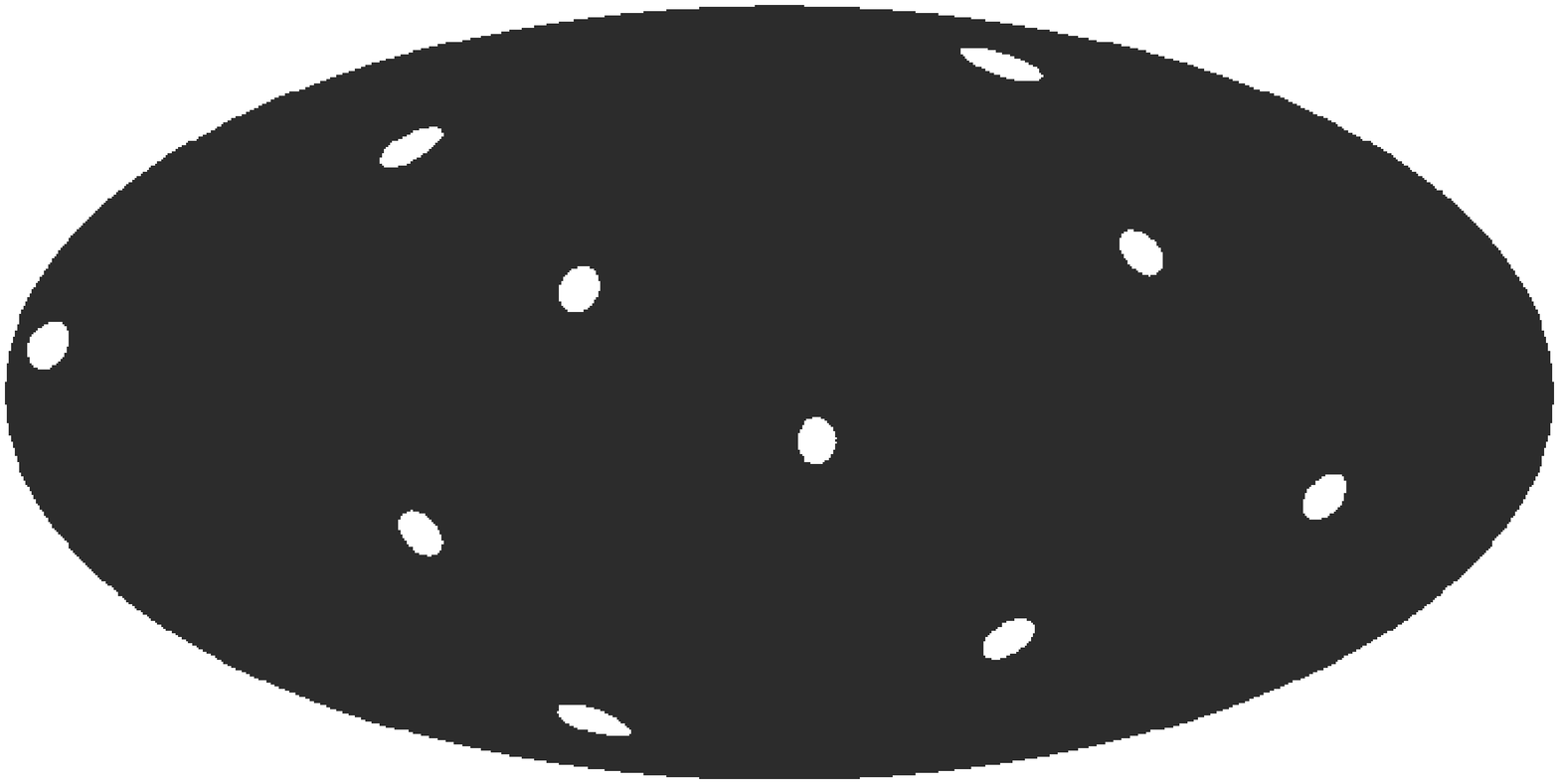,angle=90,width=4.5cm}}
\center{$\ell=5$}
\end{minipage}%
\hfill
\begin{minipage}{55mm}
\centerline{\psfig{file=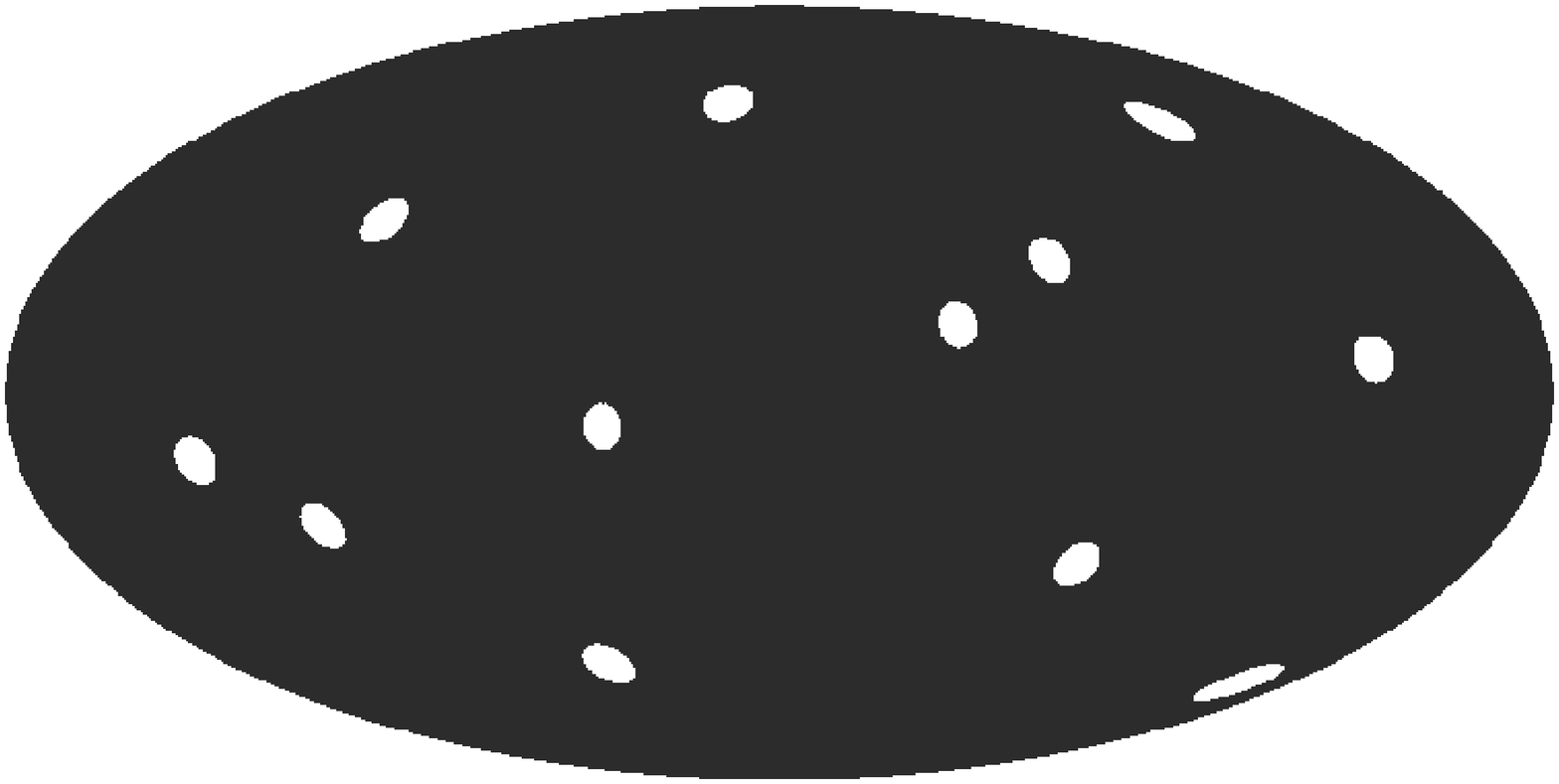,angle=90,width=4.5cm}}
\center{$\ell=6$}
\end{minipage}%
\hfill
\begin{minipage}{55mm}
\centerline{\psfig{file=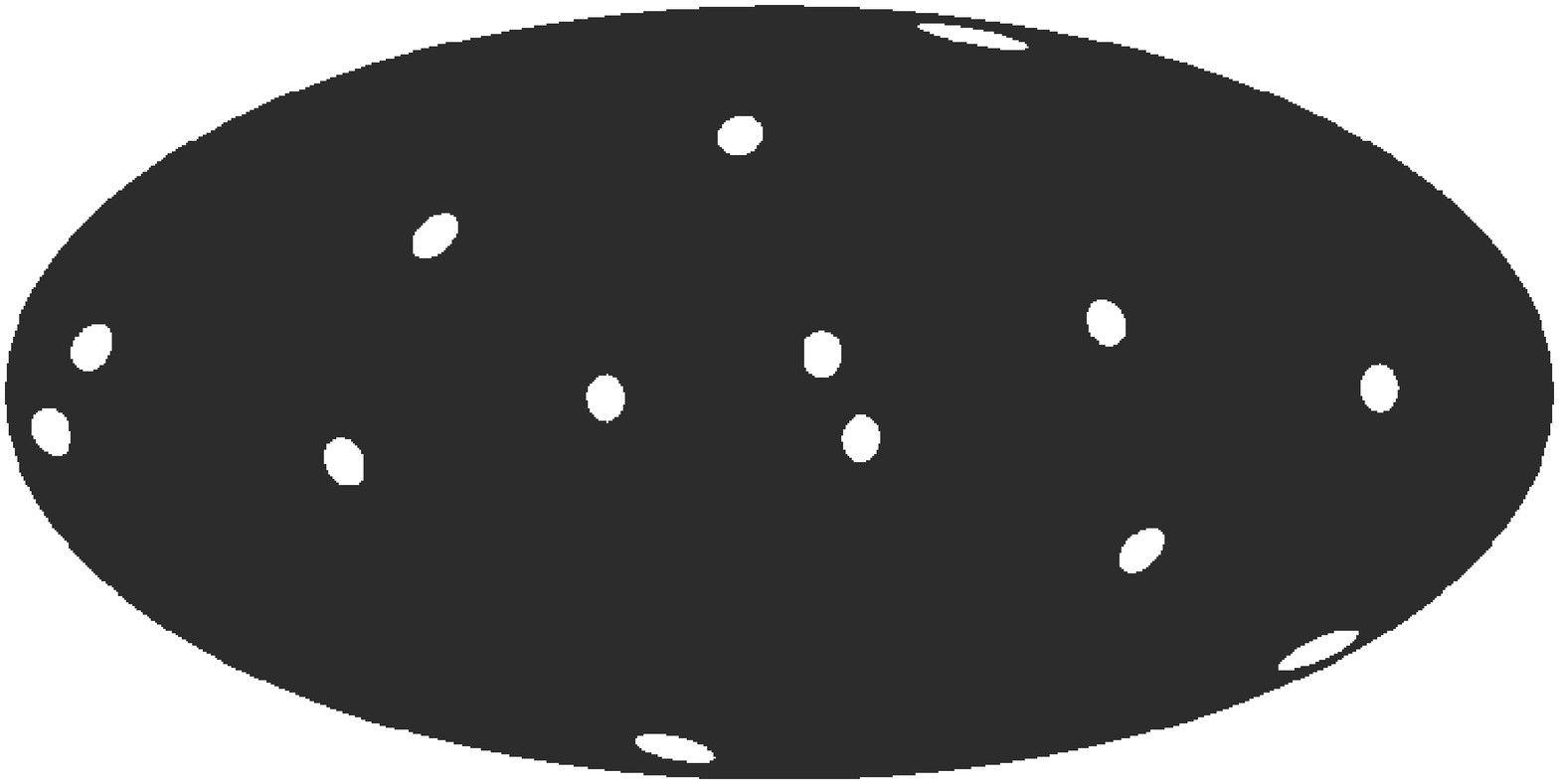,angle=90,width=4.5cm}}
\center{$\ell=7$}
\end{minipage}%
\hfill
\vspace{6mm}
\hfill
\begin{minipage}{55mm}
\centerline{\psfig{file=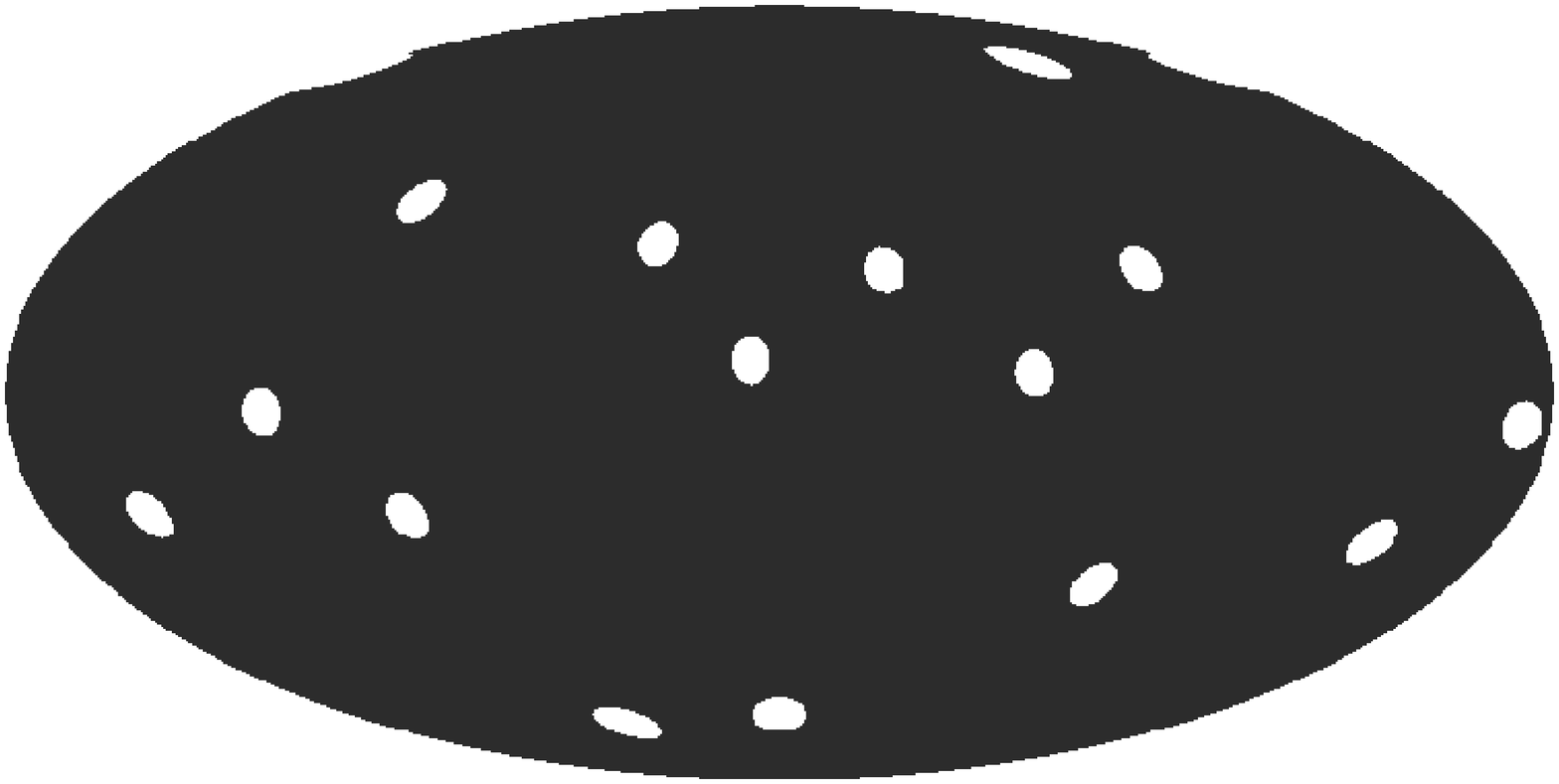,angle=90,width=4.5cm}}
\center{$\ell=8$}
\end{minipage}%
\hfill
\begin{minipage}{55mm}
\centerline{\psfig{file=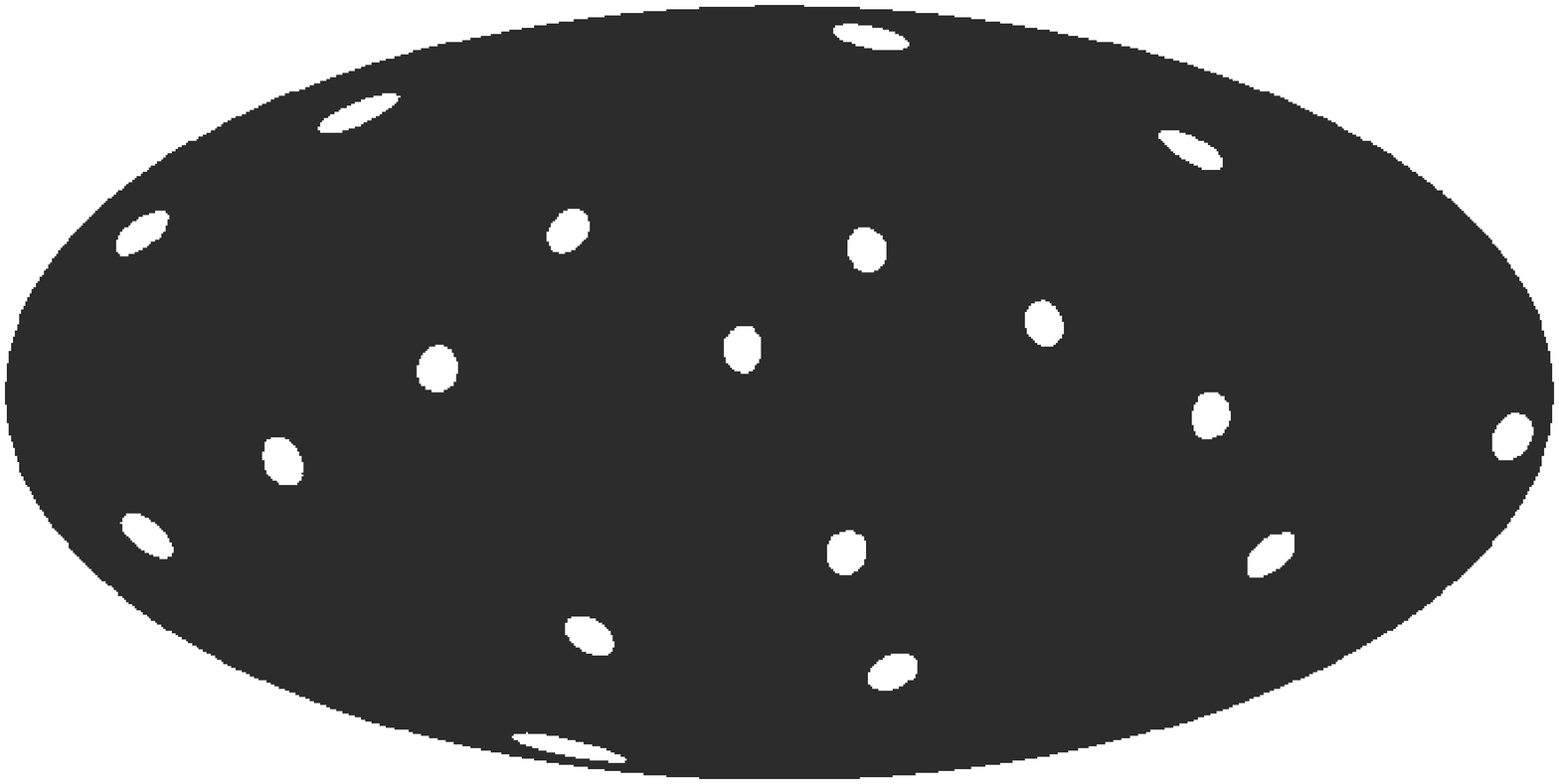,angle=90,width=4.5cm}}
\center{$\ell=9$}
\end{minipage}%
\hfill
\begin{minipage}{55mm}
\centerline{\psfig{file=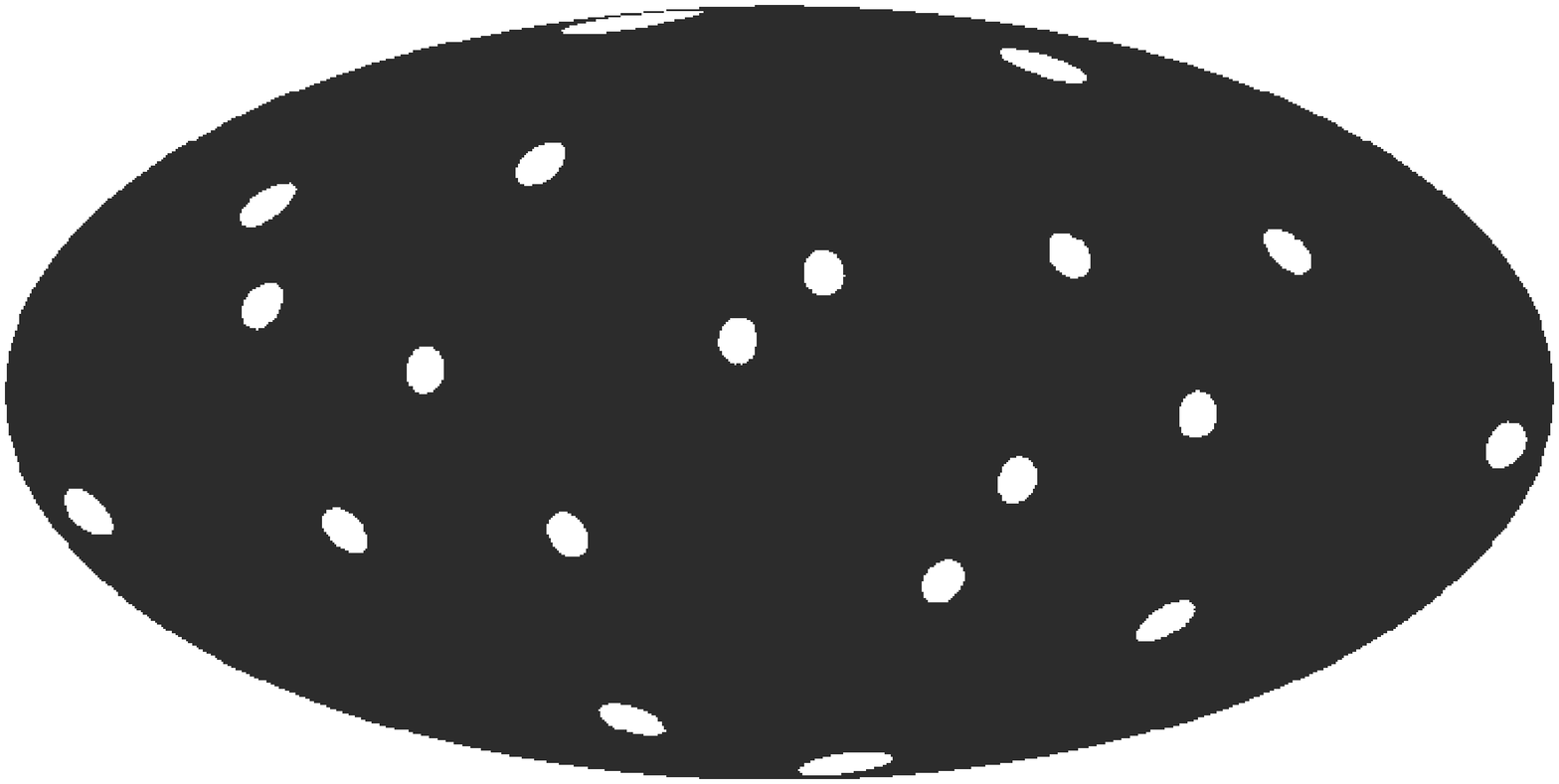,angle=90,width=4.5cm}}
\center{$\ell=10$}
\end{minipage}%
\hfill
\caption{
The Multipole Vectors for multipoles $\ell=2-10$ of the Kp0 masked 
Inverse-Noise-Squared coadded WMAP map, {\bf coadd}. The vectors are headless, so each 
vector is represented by two white dots, where it intersects with the unit sphere.}
\label{coaddvec}
\end{figure*}
\hfill
\begin{figure*}
\vspace{11mm}
\hfill
\begin{minipage}{55mm}
\centerline{\psfig{file=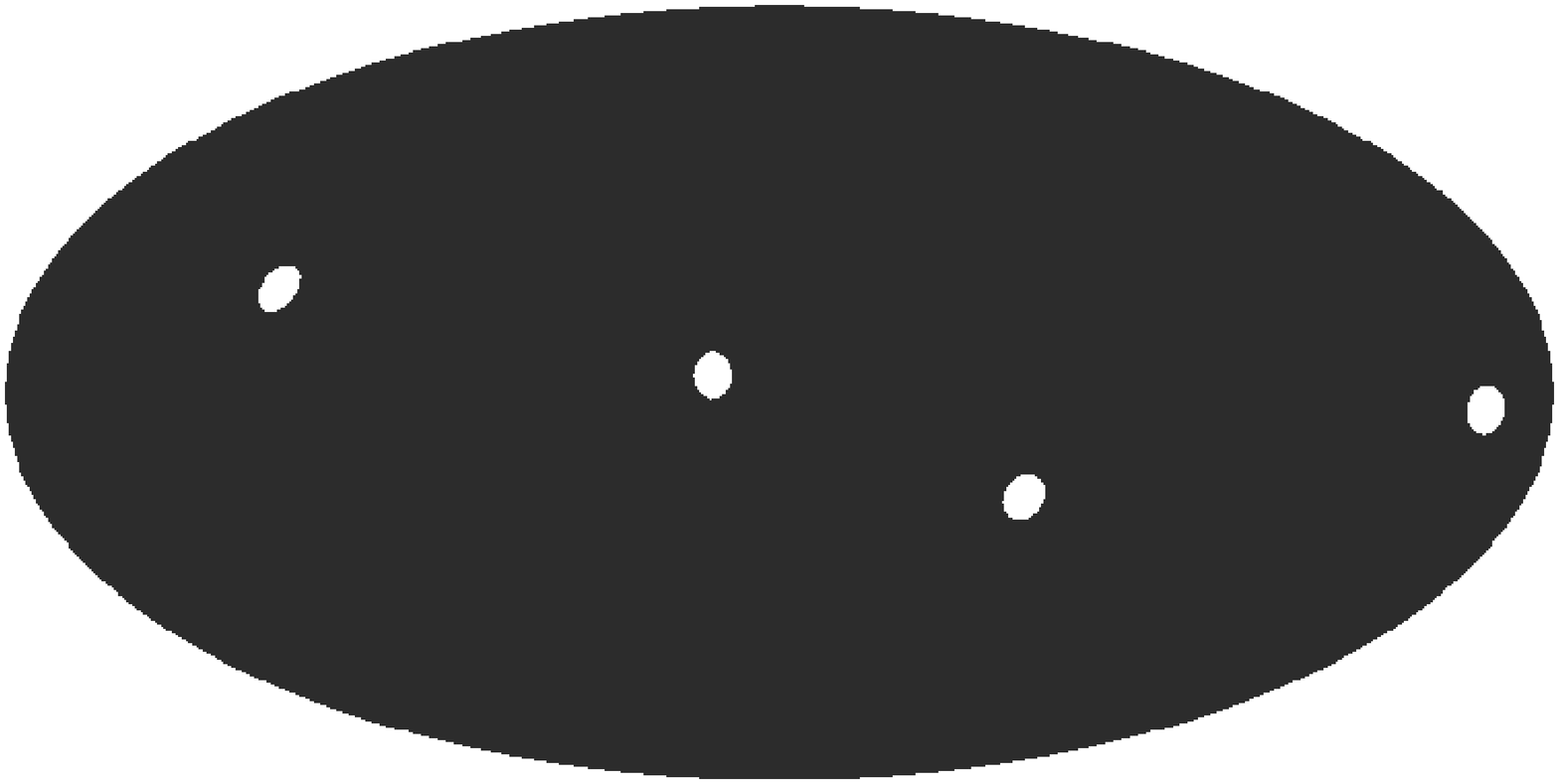,angle=90,width=4.5cm}}
\center{$\ell=2$}
\end{minipage}%
\hfill
\begin{minipage}{55mm}
\centerline{\psfig{file=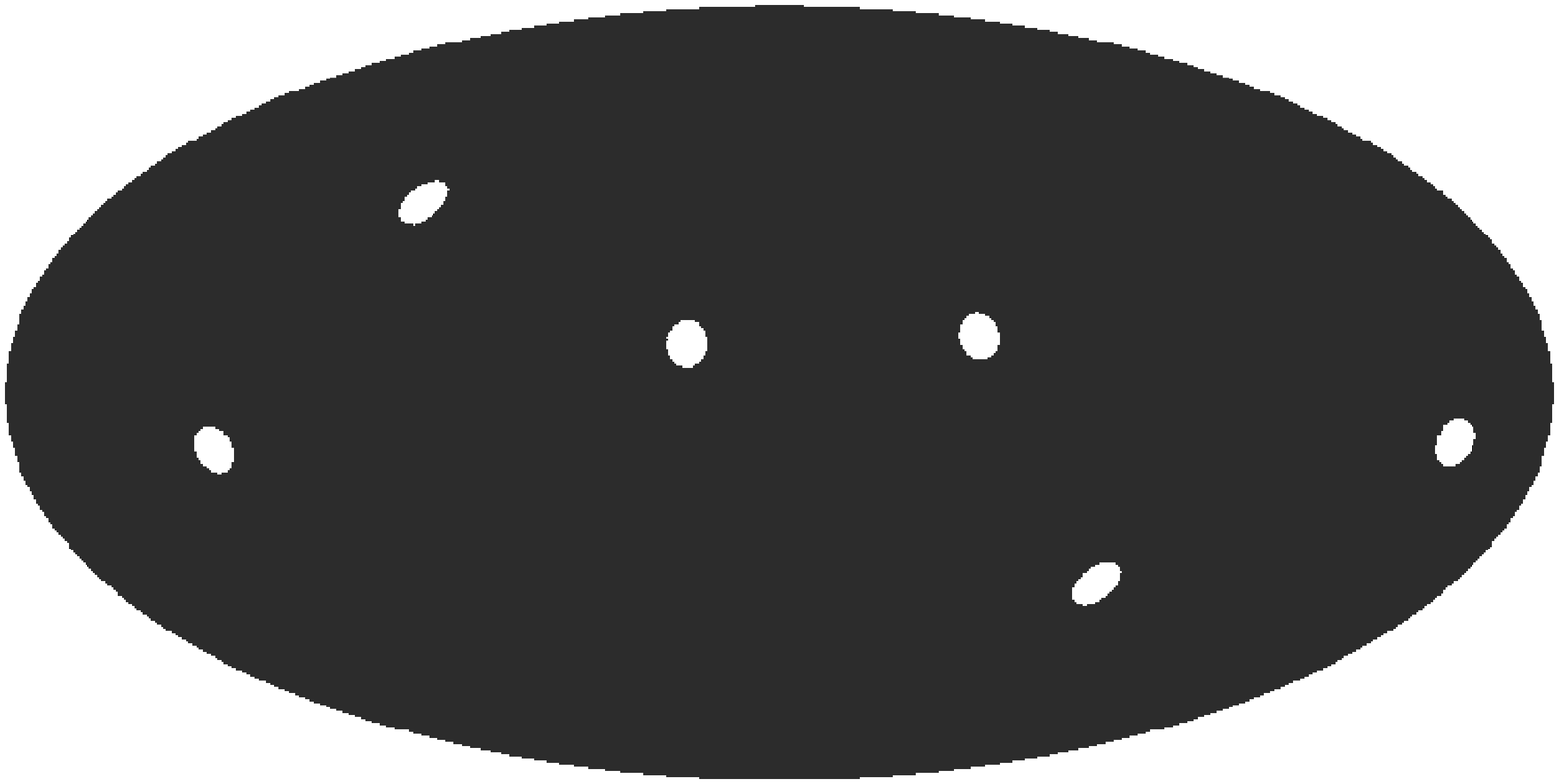,angle=90,width=4.5cm}}
\center{$\ell=3$}
\end{minipage}%
\hfill
\begin{minipage}{55mm}
\centerline{\psfig{file=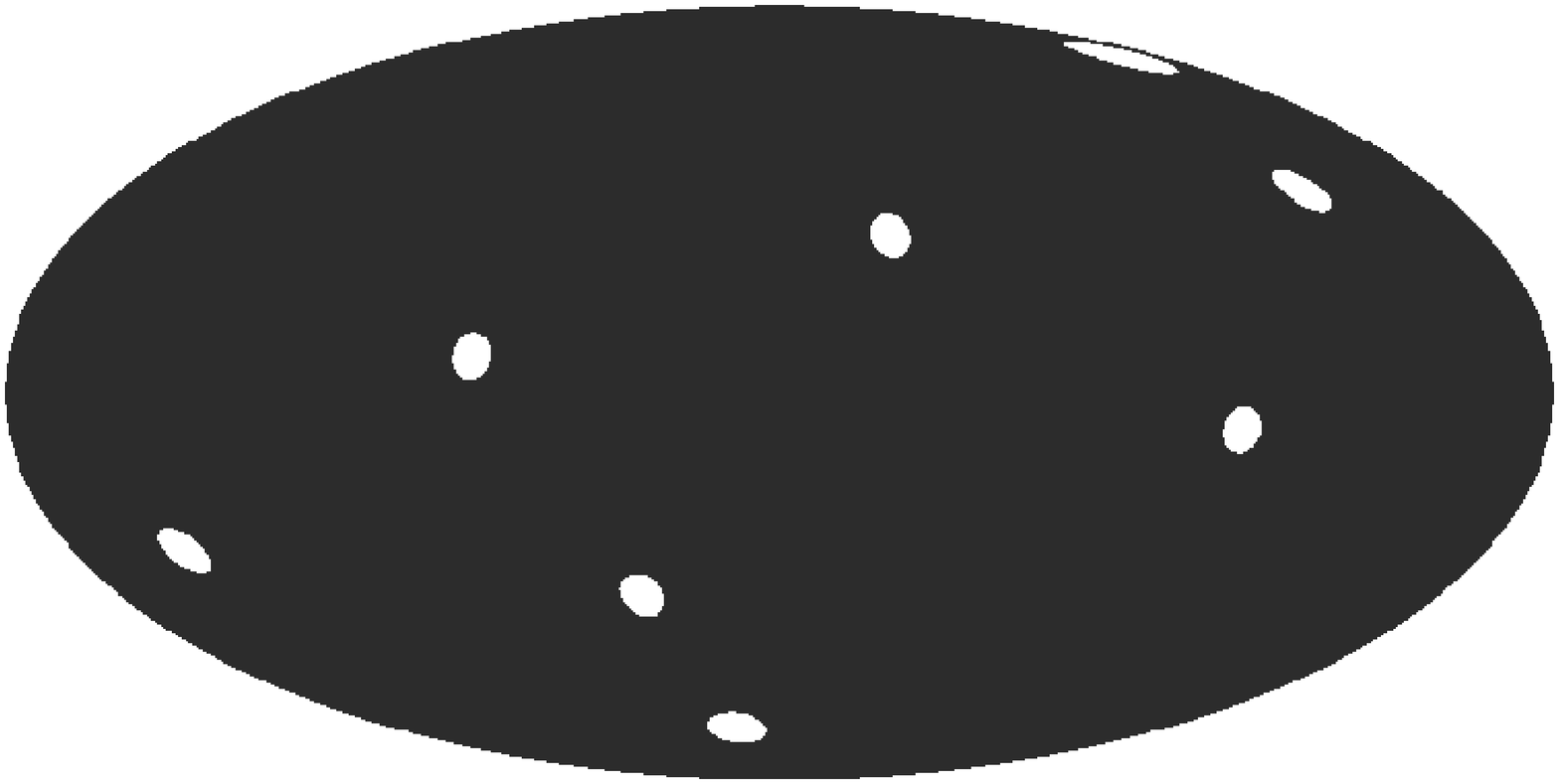,angle=90,width=4.5cm}}
\center{$\ell=4$}
\end{minipage}%
\hfill
\vspace{6mm}
\hfill
\begin{minipage}{55mm}
\centerline{\psfig{file=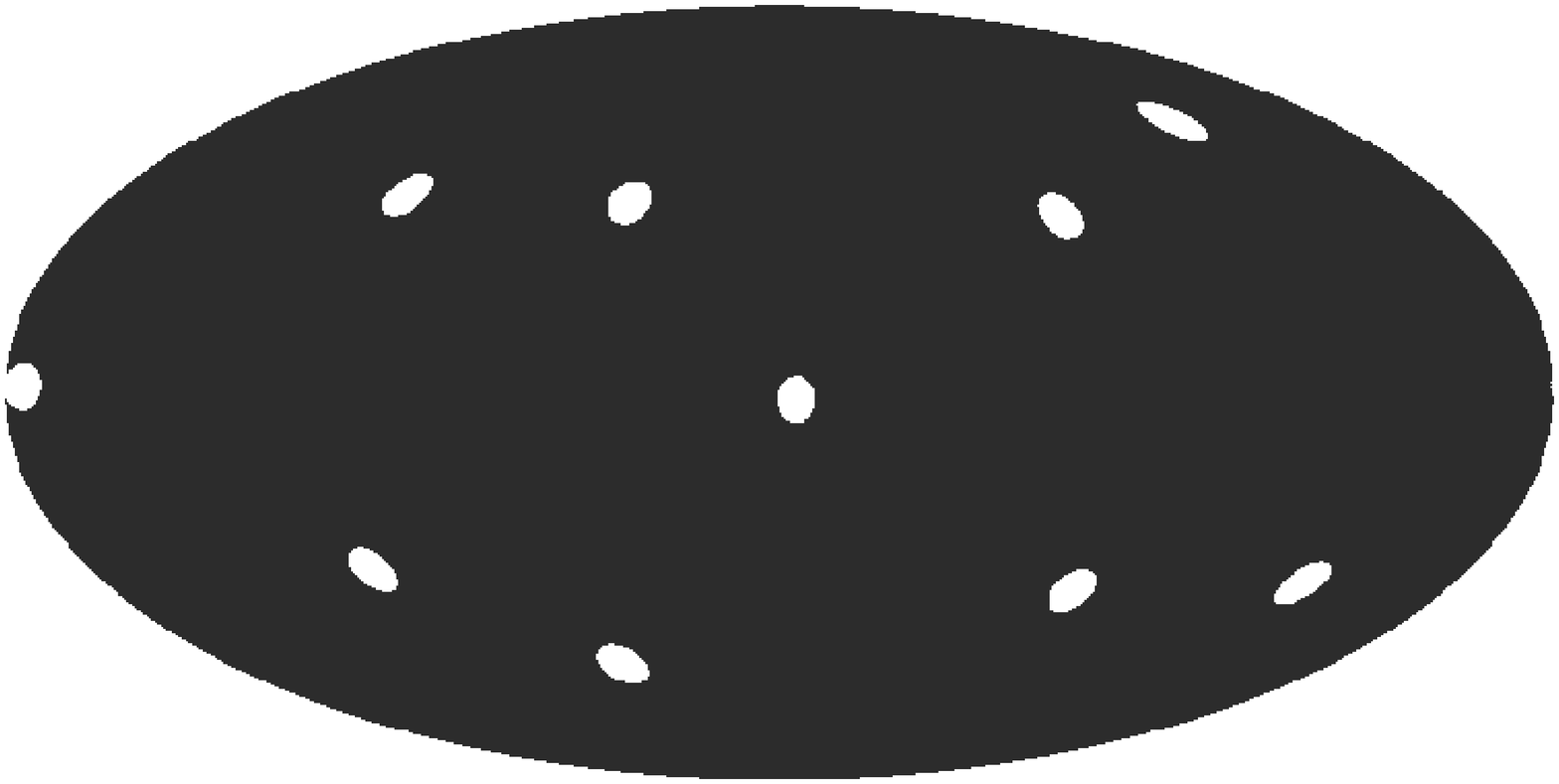,angle=90,width=4.5cm}}
\center{$\ell=5$}
\end{minipage}%
\hfill
\begin{minipage}{55mm}
\centerline{\psfig{file=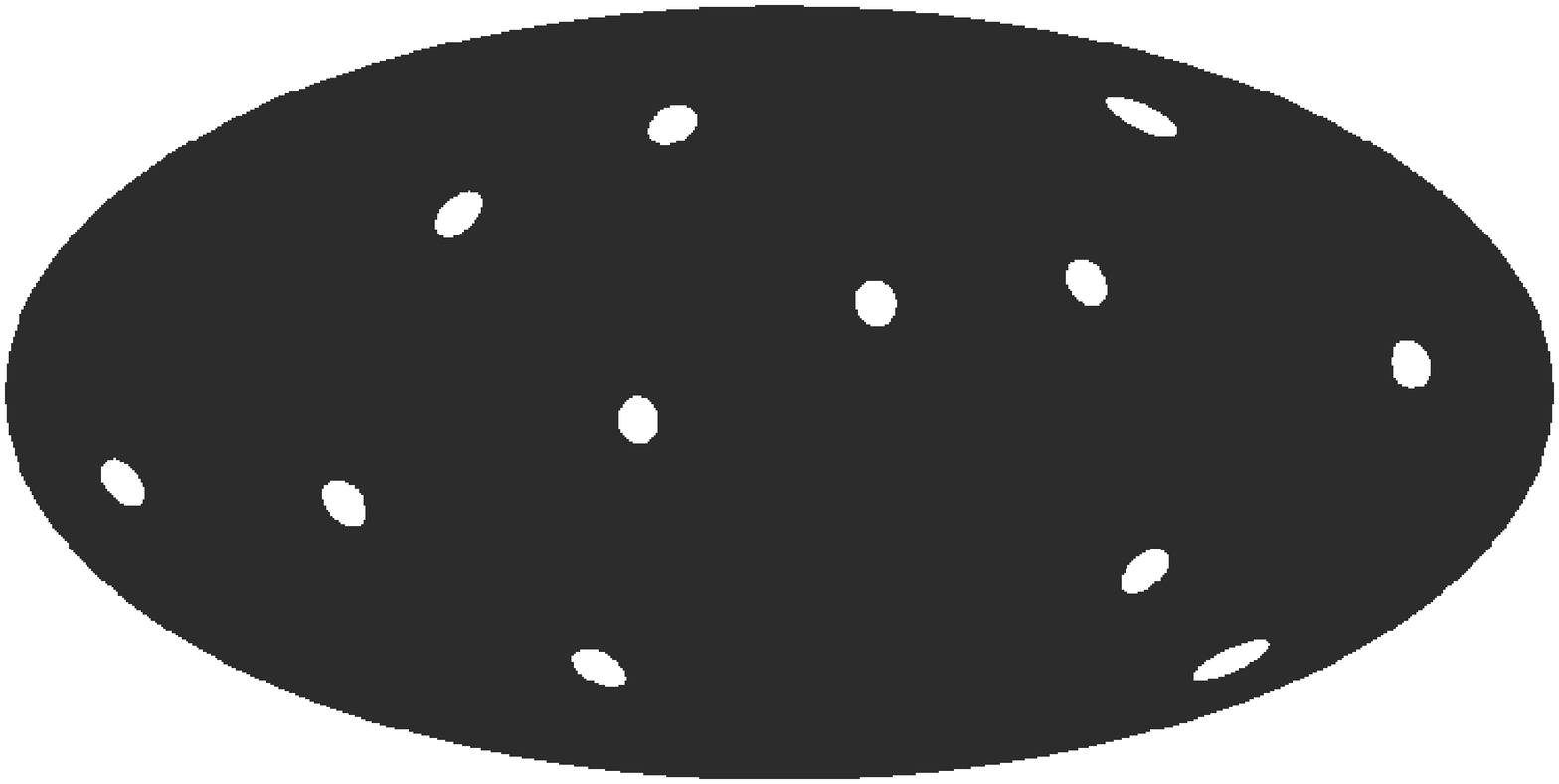,angle=90,width=4.5cm}}
\center{$\ell=6$}
\end{minipage}%
\hfill
\begin{minipage}{55mm}
\centerline{\psfig{file=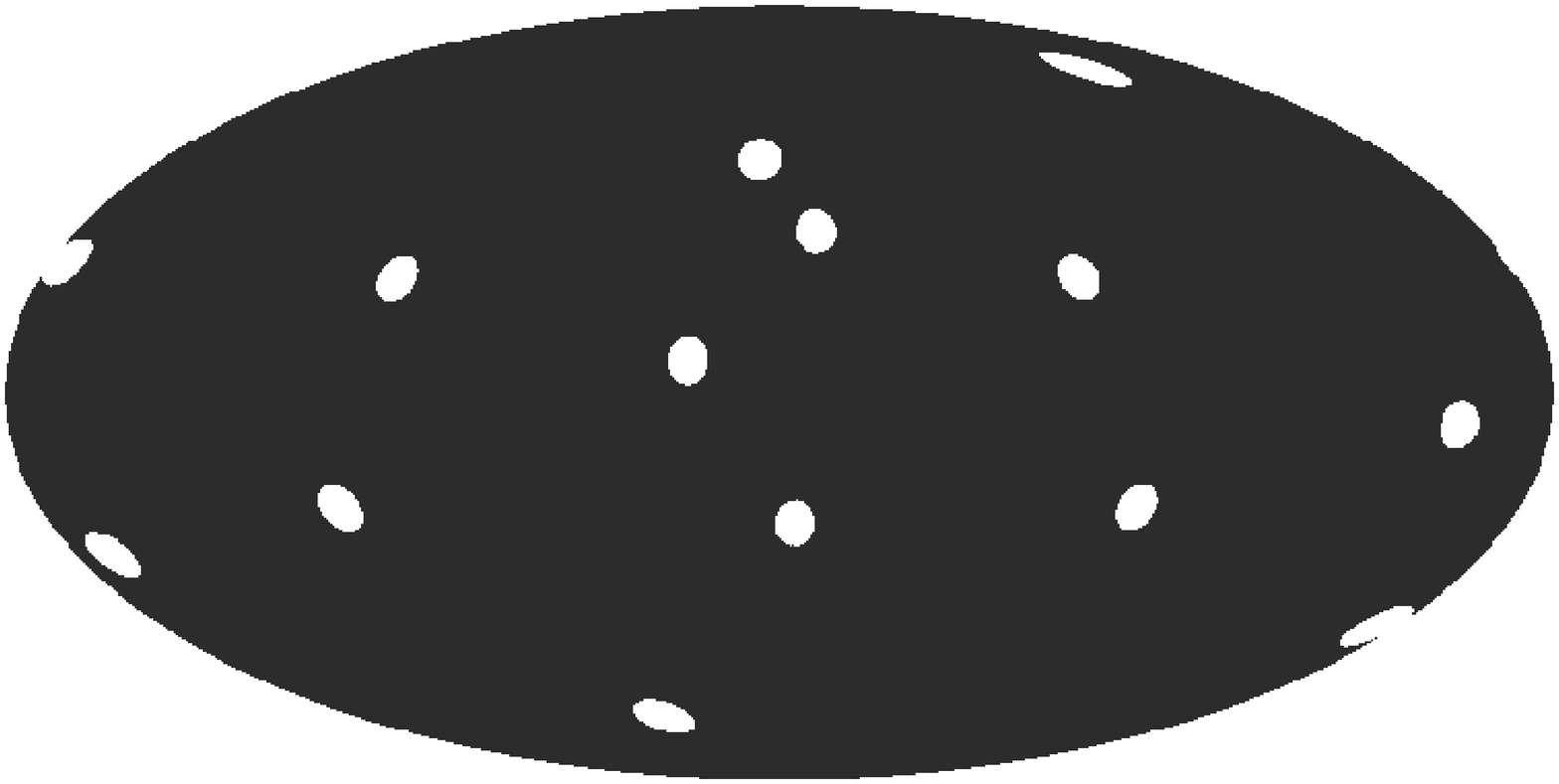,angle=90,width=4.5cm}}
\center{$\ell=7$}
\end{minipage}%
\hfill
\vspace{6mm}
\hfill
\begin{minipage}{55mm}
\centerline{\psfig{file=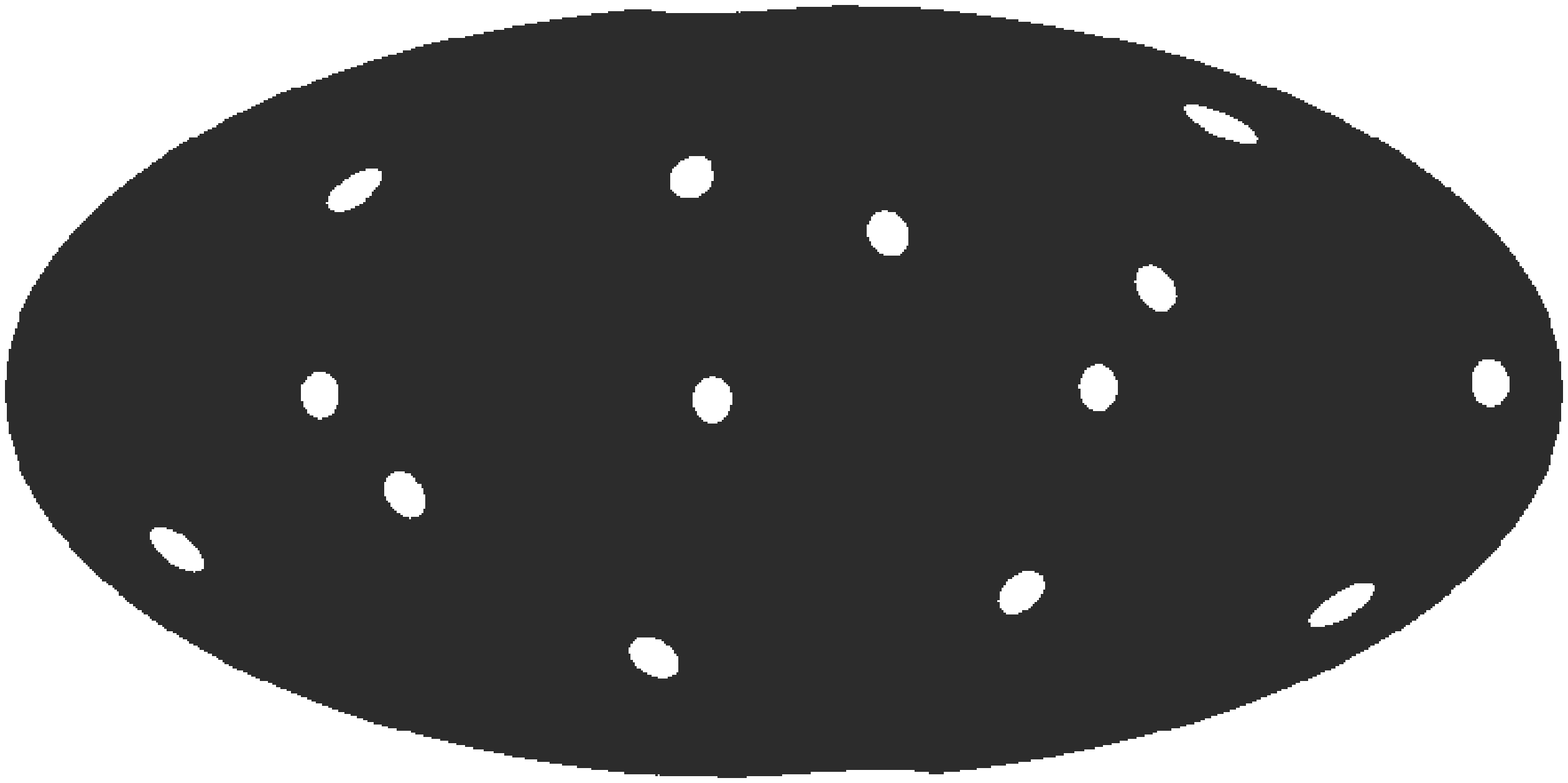,angle=90,width=4.5cm}}
\center{$\ell=8$}
\end{minipage}%
\hfill
\begin{minipage}{55mm}
\centerline{\psfig{file=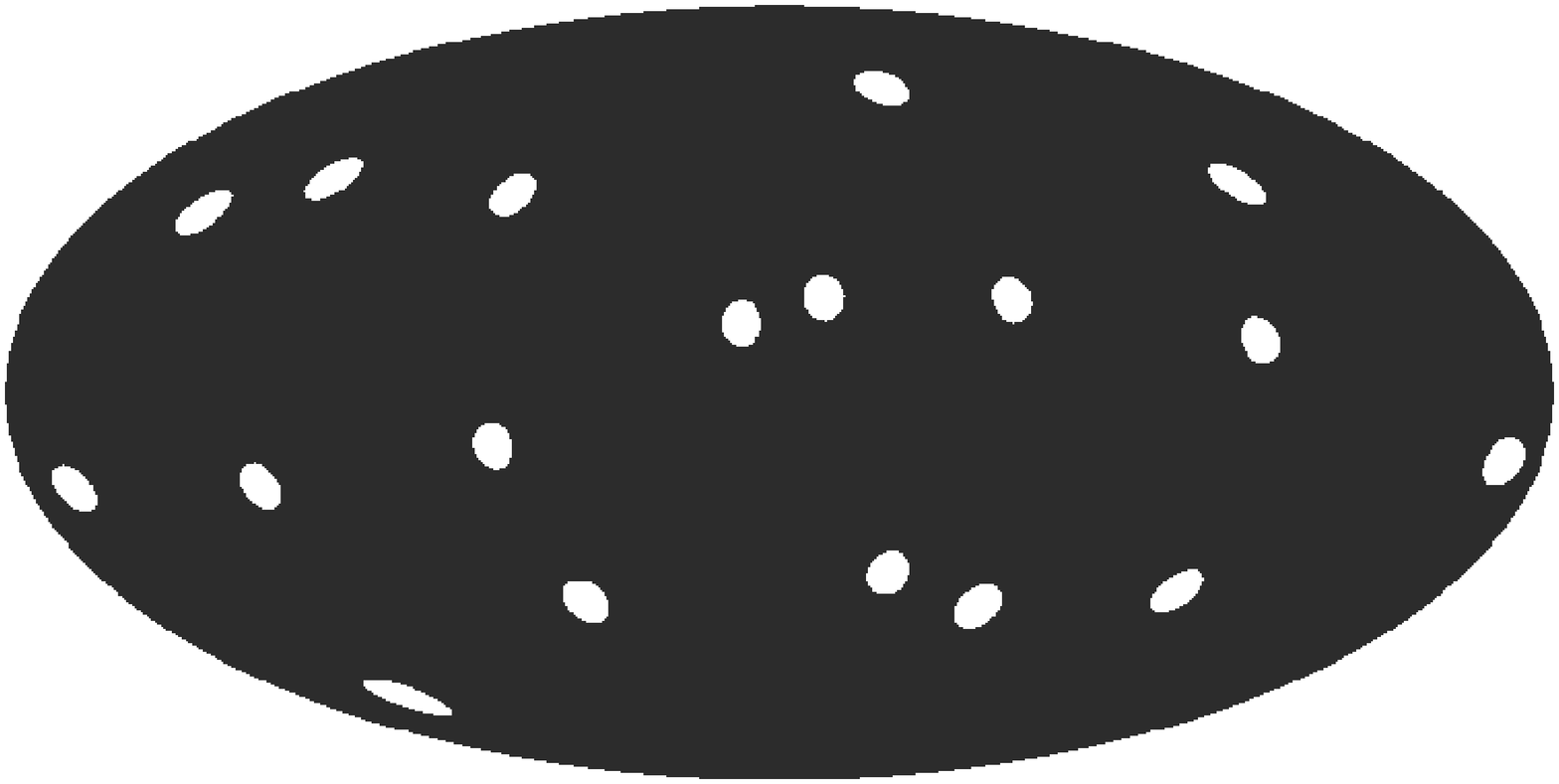,angle=90,width=4.5cm}}
\center{$\ell=9$}
\end{minipage}%
\hfill
\begin{minipage}{55mm}
\centerline{\psfig{file=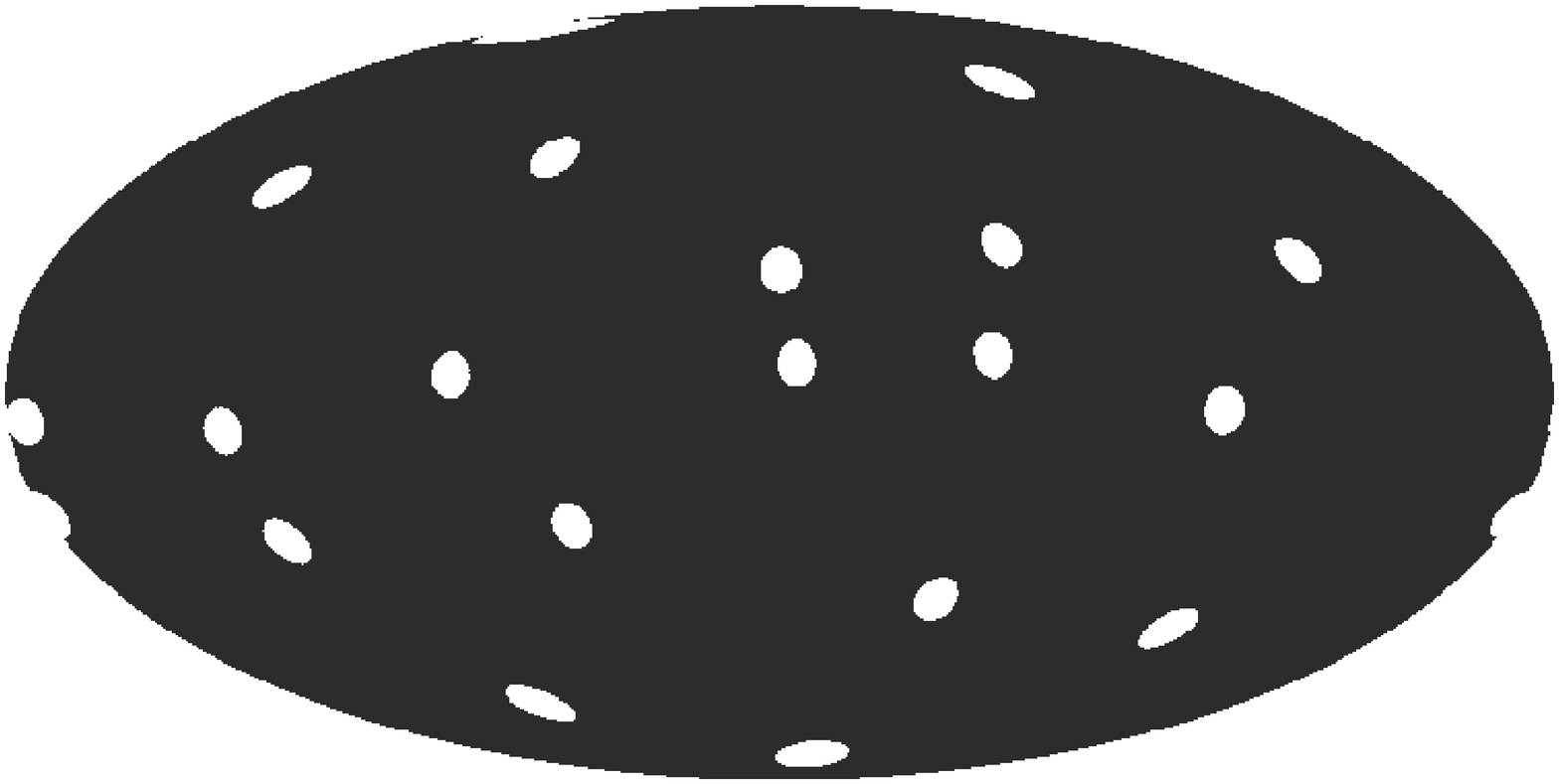,angle=90,width=4.5cm}}
\center{$\ell=10$}
\end{minipage}%
\hfill
\caption{The Multipole Vectors for multipoles $\ell=2-10$ of the 
ILC WMAP map. The vectors are headless and each 
vector is represented by two white dots, where it intersects with the unit sphere.}
\label{ILCvec}
\end{figure*}

\section{Results}\label{results}

\subsection{The datasets and methodology}

We investigate the multipole vectors of the CMB using the WMAP data. We 
look at different representations of the data in the form of 
four different maps:

\noindent 1. An Inverse-Noise-Squared Coadded map ({\bf coadd}). 
We take the 8 foreground 
cleaned maps in the Q, V, and W bands, provided by WMAP\footnote{Available at 
http://lambda.gsfc.nasa.gov/}. We combine these 8 data arrays ($DA_i$) using 
coefficients that maximally reduce the noise in the final map:
\be
T(n) =
\frac{\sum^{8}_{i=1} T_{i}(n) /
\sigma^{2}_{i}}{\sum^{10}_{i=3} 1 / 
\sigma^{2}_{i}}
\end{equation}
where $T_{i}(n)$ is the sky map for the DA $i$ with the foreground
galactic signal subtracted, and 
\bea
\sigma^{2}_{i}&=&{\sum^{Npix}_{n=1}1/\sigma^{2}_{i}(n)}\\
\sigma^{2}_{i}(n)&=&\sigma^{2}_{0,i}/Nobs_{i}(n)
\eea
$\sigma^{2}_{0,i}$ is the noise per observation for DA i, 
whose values are given by \cite{bennetta}. 
We apply the Kp0 mask to this map \citep{bennettb} 
to cut possible remaining galactic contamination near the disk,
 and then we remove any residual monopole.

\noindent 2. The Inter-Linear-Combination map ({\bf ILC}), 
produced by the WMAP team \citep{bennettb}. 
This map is formed from a combination of smoothed maps, with weights chosen 
to maintain CMB anisotropy signal while minimizing the Galactic foreground. The 
WMAP team advise that the {\bf ILC} map be used as a visual tool, and not used for 
CMB studies. We include it here for interest, as other groups have claimed 
interesting detections of anomalies in this map\citep{copi1}.

\noindent 3+4. \cite{teg} produced their own foreground cleaned maps
\footnote{Available at www.hep.upenn.edu/~max/wmap.html}. Their approach did not 
assume a specific power spectrum, and these maps contain less foreground signals and 
noise than the ILC map. For the lowest multipoles the maps are clean enough that 
no galactic cut is required~\citep{teg}. Our third map is their cleaned map 
({\bf TOHcl}), and our fourth map is their Wiener filtered map ({\bf TOHw}).

We will compare findings to those from simulations. Our simulations are of 
Gaussian fluctuations, with the noise and beam properties of the 
{\bf coadd} map. We perform 1000 of these simulations with the Kp0 mask, and 
1000 without. In the case of the {\bf coadd} map we always compare the results 
to those with the Kp0 mask added, and for the other {\bf ILC}, {\bf TOHcl}, 
and {\bf TOHw} maps we compare their results to those without the mask. We 
have only done simulations with the noise 
and beam properties of the {\bf coadd} map to simplify matters. We find this 
acceptable as at the low-$\ell$ multipoles we are concentrating on, 
the difference in the levels of noise and resolution seen in the four maps 
are insignificant.

Throughout our analysis we use software produced by the HEALPix collaboration
, \cite{healp}\footnote{Available at http://www.eso.org/science/healpix}. 
To compute the Multipole Vectors we use codes produced by 
\cite{copi1}\footnote{Available at www.phys.cwru.edu/projects/mpvectors/}. 

\subsection{The Multipole vectors}
In Figures~\ref{coaddvec},\ref{ILCvec} we display in Mollweide projection 
the multipole vectors for map {\bf coadd} and 
{\bf ILC}, for multipoles $\ell=2-10$ (for these multipoles the vectors for 
maps {\bf TOHcl} and {\bf TOHw} by eye look very similar to the {\bf ILC} 
map results). 
Recall that these vectors 
are headless, so we show dots at both ends. In Figure~\ref{ILCvec} one can see 
the planarity of the $\ell=3$ multipole vectors, and the alignment of this plane 
with the galactic plane and the plane of the $\ell=2$ multipole vectors, as 
reported by \cite{oliv,copi1,copi2,weeks}. However, a word of warning, this 
visual projection can be misleading as other significant features would not be 
so apparent by eye. In this projection our eye easily picks out any planarity in 
the galactic plane.

We now use our prescription of defining ``anchor vectors'' to investigate the 
multipole-frames and the multipole-dots for each map.
Therefore the analysis of CMB data now splits into two areas:

\noindent 1. The Multipole Frames. We search for inter-$\ell$ 
correlations by looking for correlations between these frames. SI demands that the 
frames be uniformly orientated (apart from the effect of sky cuts, noise, etc).
Any correlation between frames of different multipoles, apart from that seen 
in the simulations, would indicate statistical 
anisotropy. The simplest case would be an alignment of the frames - clearly 
indicating one preferred frame. We investigate this issue for our four maps
 in Section~\ref{frames}.

\noindent 2. The Multipole Invariants. 
For each multipole we extract the $2\ell-3$ invariants, $X_{ij}$. 
From simulations we can extract their expected distributions 
assuming SI and Gaussianity taking into account additional 
factors such as noise and beam. 
We investigate this in Section~\ref{dots}.

\subsection{The Multipole Frames}\label{frames}
For each of the four maps we do the following. For each multipole in the range
$\ell=2-20$ we find the Multipole Vectors, 
and order them as described in Section~\ref{anchors}. We define $L_1$ and $L_2$ 
as the anchor vectors, and use these anchor vectors to define an orthonormal 
frame as described in Section~\ref{multvec}.

\begin{figure*}
\begin{minipage}{55mm}
\centerline{\psfig{file=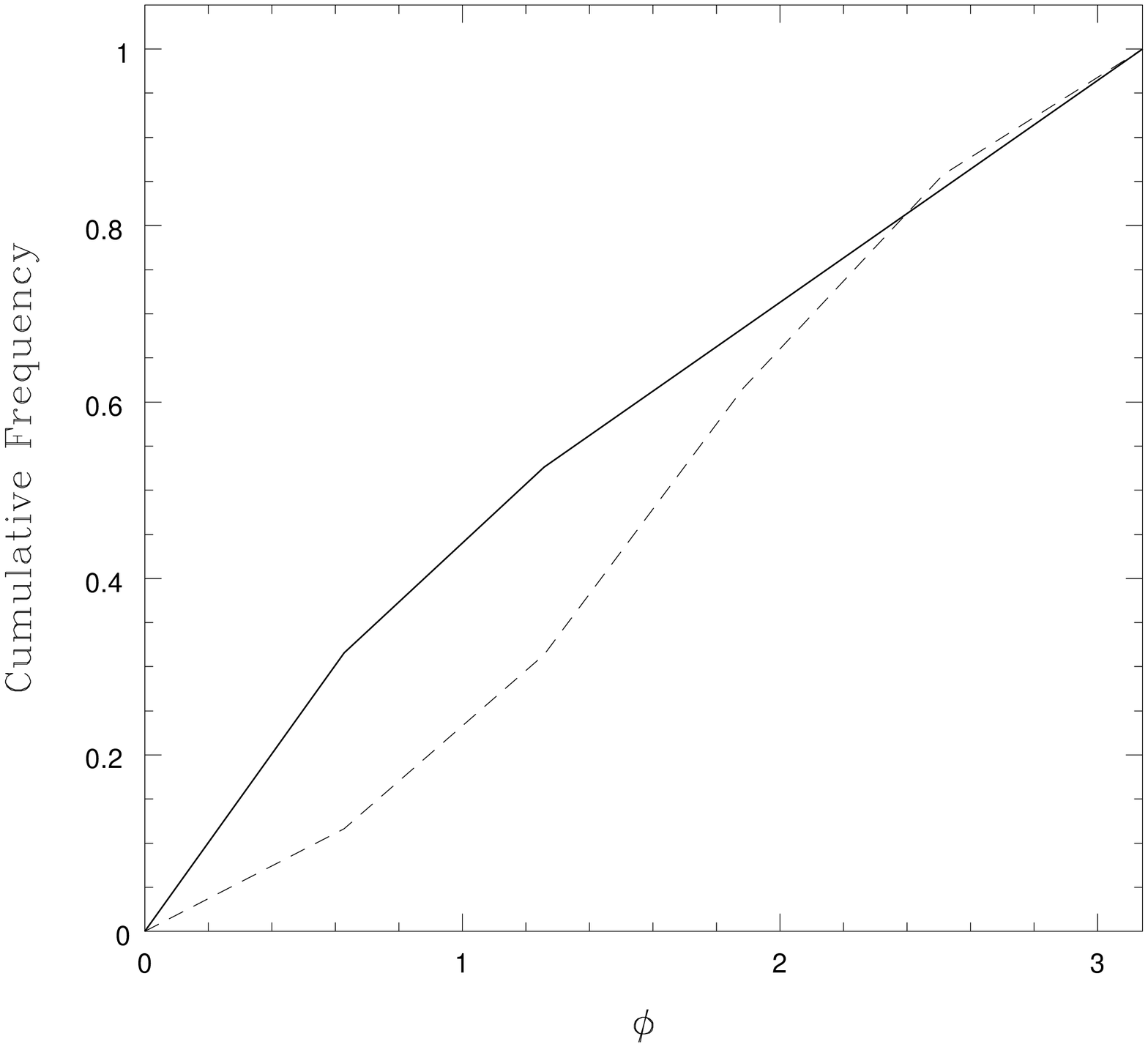,width=4.5cm}}
\end{minipage}%
\hfill
\begin{minipage}{55mm}
\centerline{\psfig{file=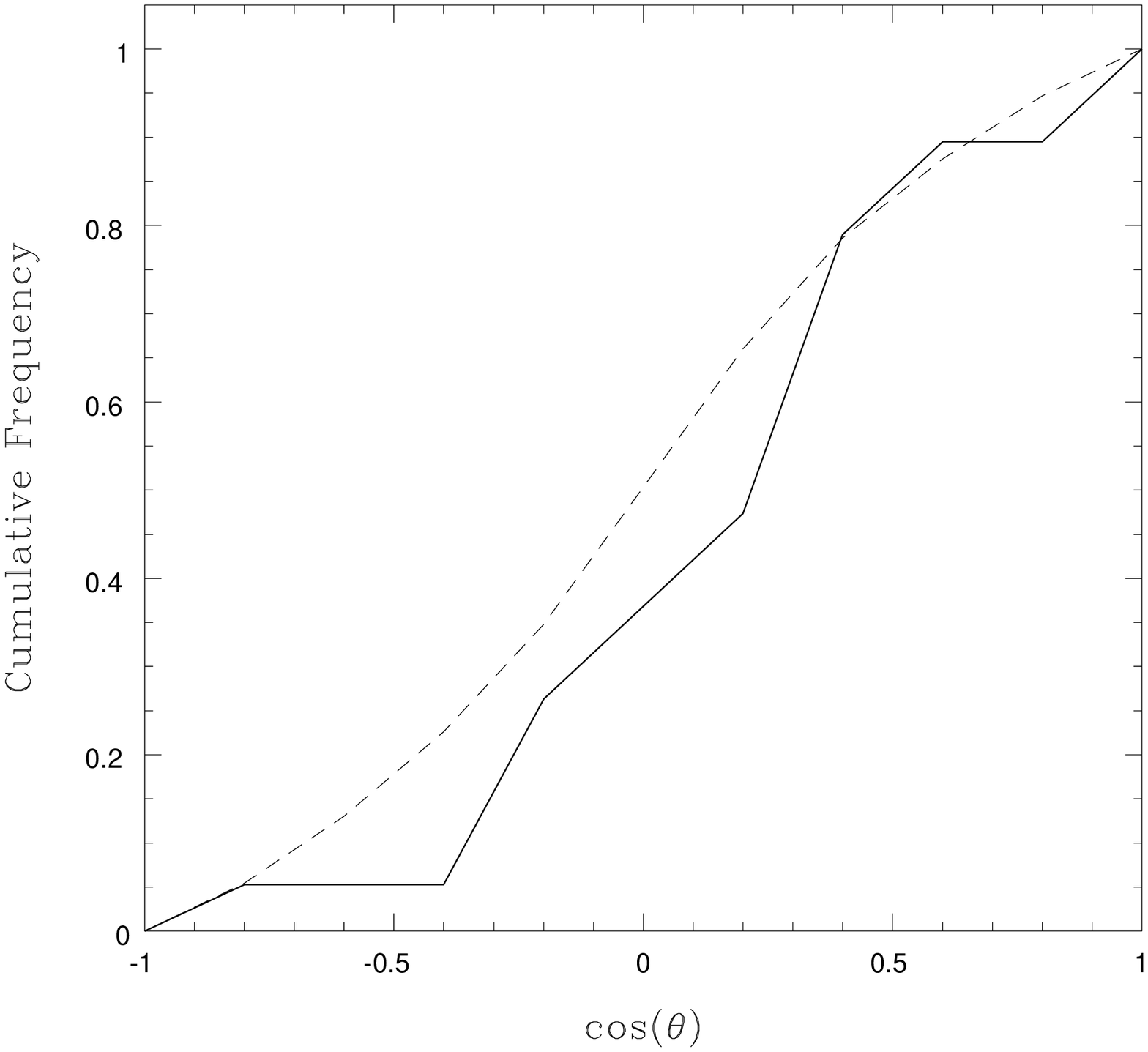,width=4.5cm}}
\end{minipage}%
\hfill
\begin{minipage}{55mm}
\centerline{\psfig{file=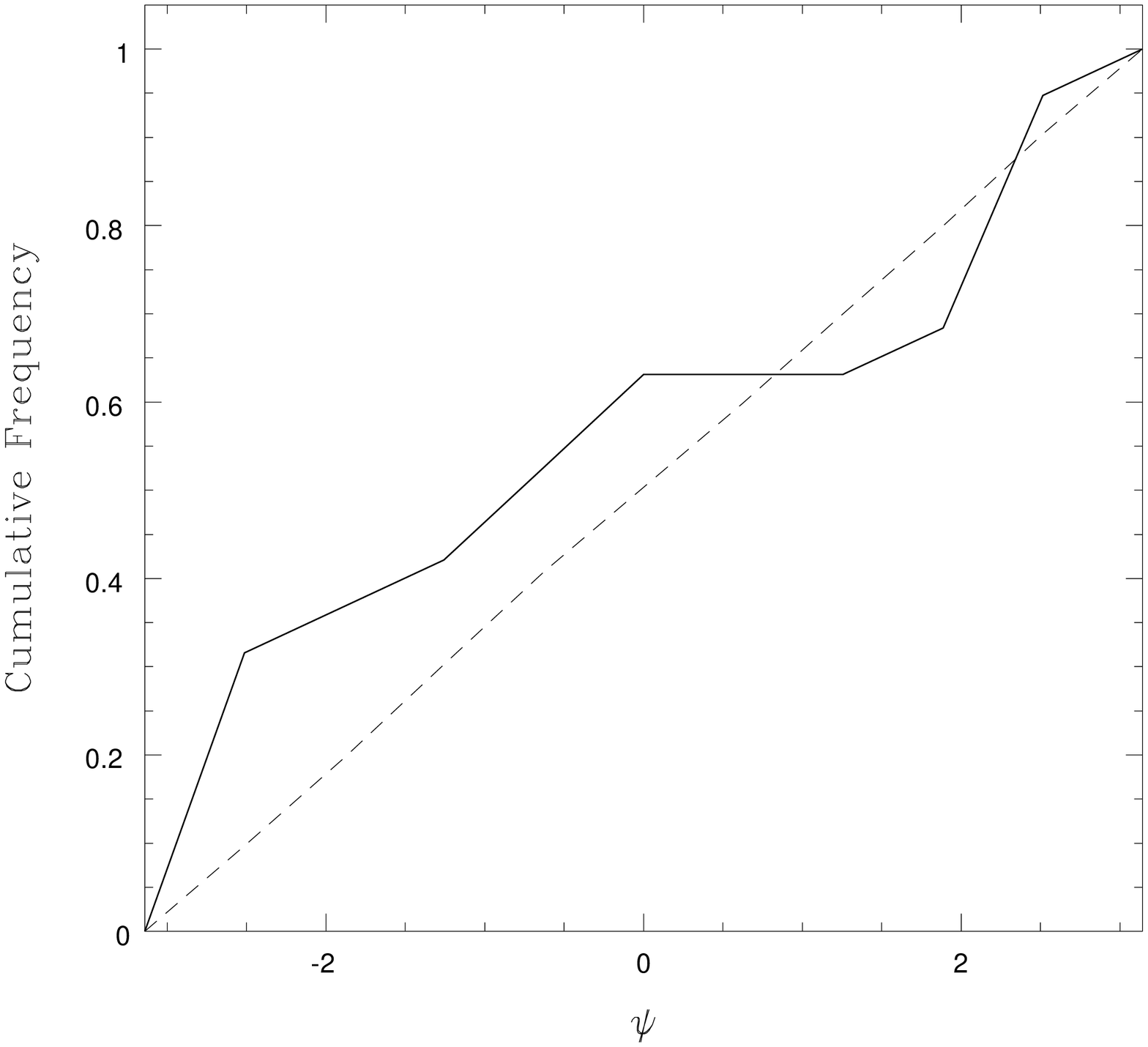,width=4.5cm}}
\end{minipage}%
\hfill
\caption{The cumulative frequency plot of the Euler angles from the multipole frames 
of the Coadded WMAP map, for $\ell=2-20$. Dashed line= the corresponding 
expected distribution returned from simulations.}
\label{coadded}
\end{figure*}

Under the hypothesis of Statistical Isotropy we expect these multipole frames 
to be uniformly distributed (except for the anisotropy induced by the galactic cut in 
the case of the {\bf coadd} map).
One cannot assess how uniformly distributed a frame is one at a time as 
clearly every orientation is just as likely and so no significance can be 
assigned to any particular result. One can 
only ask how probable it is that a number of frames came from a uniform distribution. 
To do this we use a Kolmogorov-Smirnov test. This provides a goodness of fit test 
for a statistical distribution. It looks at the discrepancy between the 
cumulative frequency of observations and the expected cumulative frequency. The 
maximum discrepancy is called the D-statistic. For large enough sample sizes there 
is an analytical expression for the critical values of the D-statistic, over which 
a null hypothesis should be rejected.

We calculate the cumulative frequency over a range of multipoles of the Euler angle 
observations, $\phi, cos(\theta), \psi$.
We do this for two multipole ranges, $\ell=2-10$ and $\ell=2-20$. 
 We compare the distributions we find for the maps to those from simulations. 
As expected, the distributions the simulations return are near uniform.
 For example, in Figure~\ref{coadded} we visualise the cumulative 
frequencies of the {\bf coadd} map results, accumulated over multipoles $\ell=2-20$. 
The expected distribution returned by the simulations is also shown. The D-statistic is the 
maximum discrepancy between the two lines. 

\begin{table}
\center
\begin{tabular}{|l|ccc|}
\hline
     & $\phi$      & $cos(\theta)$ & $\psi$   \\
\hline
\hline
     &            &  $\ell=2-10$   &          \\
\hline
coadd & 0.335(9.0) & 0.205(51.5) & 0.360(12.0) \\
ILC & 0.079(92.3) & 0.256(35.5)	& 0.284(26.2) \\
TOHcl & 0.370(5.1) & 0.174(73.8) & 0.138(90.3) \\
TOHw & 0.259(25.2) & 0.145(87.7) & 0.249(40.5) \\
\hline
\hline
    &             & $\ell=2-20$ &  \\
\hline
coadd & 0.213(7.8) & 0.187(25.3) & 0.219(17.4)\\
ILC &  0.130(49.0) & 0.070(98.4) & 0.069(95.9)\\
TOHcl & 0.133(44.5) & 0.225(14.8) & 0.109(80.6)\\
TOHw & 0.133(44.5) & 0.075(94.1) & 0.136(57.1)\\
\hline
\end{tabular}
\caption{D-statistics from cumulative frequency distributions of the Multipole 
Frame Euler angles from multipoles $\ell=2-10$ and $\ell=2-20$.}
\label{D-statFRAMES}
\end{table}


For each of the four maps, and for both the $\ell$ ranges we find the 
maximum discrepancy, the D-statistic, that the 3 Euler angles find. 
As our sample sizes are relatively small (9 observations for $\ell=2-10$ range, and 19 for 
$\ell=2-20$) we do not use analytical expressions for the critical values of the 
D-statistic: we assess the significance of 
the D-statistics we observe by comparing them to all those the simulations found.
In Table~\ref{D-statFRAMES} we list the D-statistics found and, in brackets, the 
percentage of simulations that found larger values. 
A significant departure from the null hypothesis would result in a high D-statistic, and 
therefore a low percentage value. 
We find no significant evidence for any departure from the expected uniform (nearly 
uniform in the case of the {\bf coadd} map) distribution for the multipole frames. Therefore, using 
our Multipole Frames we find no evidence for a departure from SI for multipoles $\ell=2-20$.

\subsection{The Multipole Invariants}\label{dots}

We now turn our attention to the invariant degrees of freedom. As discussed in 
Section~\ref{sep}, the invariants of one multipole will probe Gaussianity, and the 
m-preference associated with SI. In terms of the Multipole Vectors this information is 
stored in their relative spacing - the angles between one and another. \cite{dennis2} 
computed the analytical expression for the expectation of the general dot product for 
a given multipole. 

For each multipole we will not look at all the dot products between all the vectors 
as this contains redundant information. We will use our prescription of defining 
anchor vectors, and observe the $2\ell-3$ invariants $X_{ij}$, as described in 
Section~\ref{sep}.

\begin{figure*}
\centering
\hfill
\begin{minipage}{55mm}
\begin{center}
\psfig{file=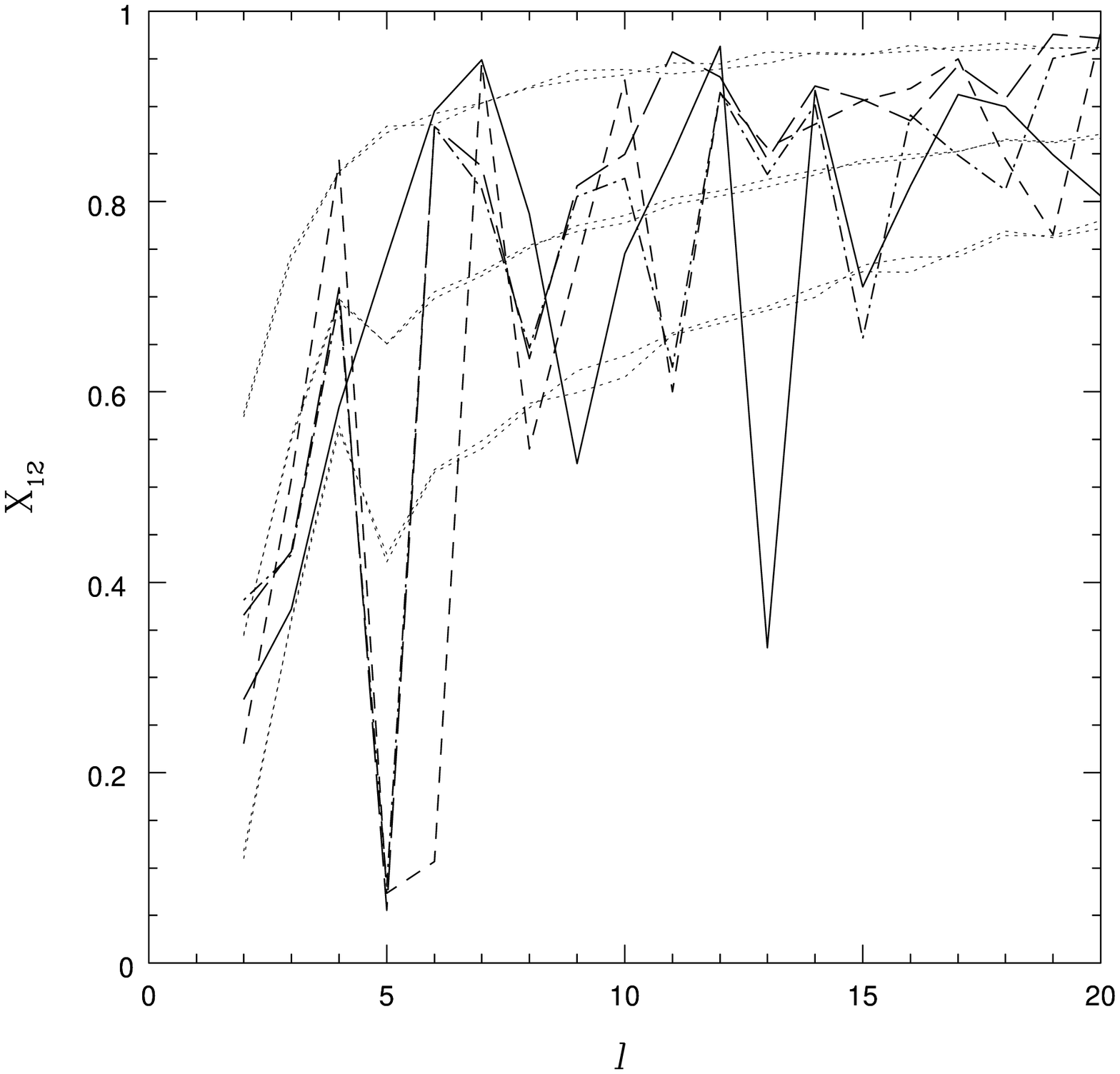,width=5.5cm}
\end{center}
\end{minipage}%
\hfill
\begin{minipage}{55mm}
\begin{center}
\psfig{file=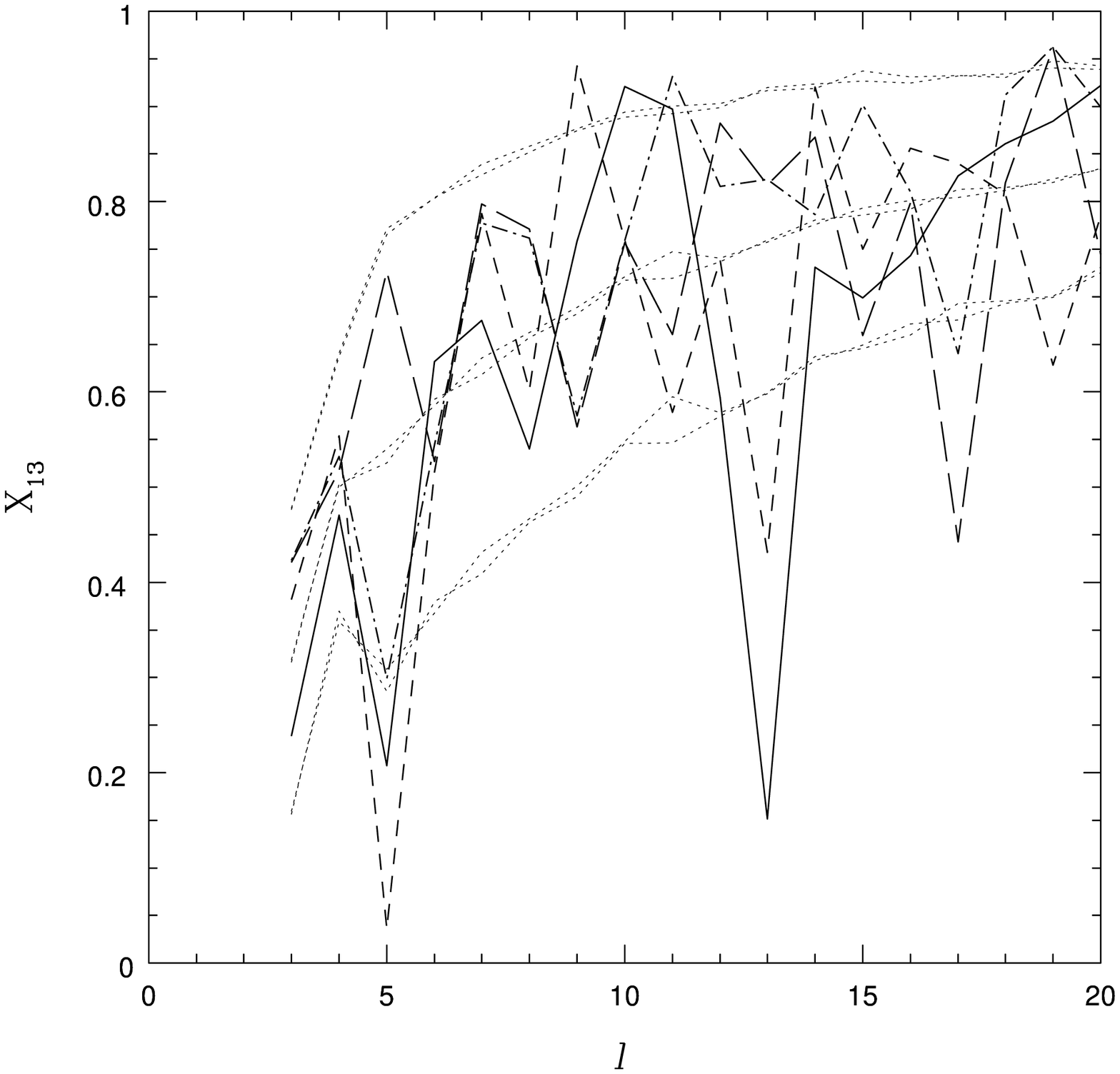,width=5.5cm}
\end{center}
\end{minipage}%
\hfill
\begin{minipage}{55mm}
\begin{center}
\psfig{file=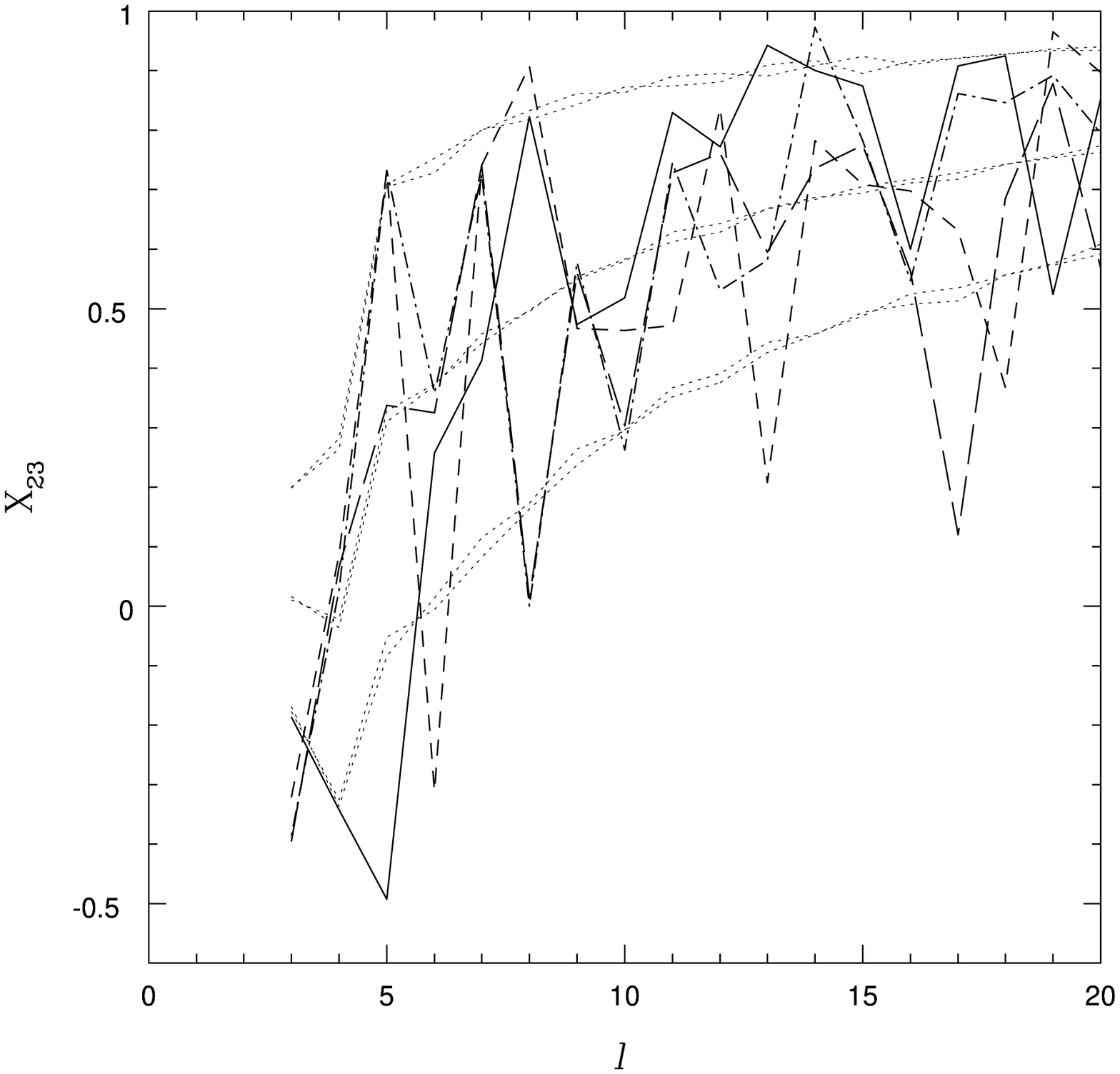,width=5.5cm}
\end{center}
\end{minipage}
\hfill
\caption{The $X_{12}$, $X_{13}$, and $X_{23}$ results for the {\bf coadd} map (bold line), the 
{\bf ILC} map (short-dash), the {\bf TOHcl} map (long-dash), and the {\bf TOHw} map (dash-dot). Also 
shown is the mean, $\pm 1 \sigma$ for the simulations (dotted), with and without the Kp0 mask added. }
\label{Xij}
\end{figure*}

For each map, and for each multipole $\ell=2-20$ we calculate the $2\ell-3$ invariants 
$X_{ij}$. We compare their values to those from simulations. 
In Figure~\ref{Xij} we display the results found for the invariants $X_{12}$, 
$X_{13}$, and $X_{23}$. We also show the $\bar{X}_{ij}\pm \sigma$ values that the 
simulations returned.

In these plots we can observe potentially anomalous features. For example in the 
$X_{12}$ plot the $\ell=5$ value is very low for the {\bf ILC}, {\bf TOHcl}, and 
{\bf TOHw} maps, and the $\ell=13$ value is low for the {\bf coadd} map. 
The $X_{13}$ plot again shows out-lier values for $\ell-5$ in the {\bf ILC} map, and 
$\ell=13$ for the {\bf coadd} map.

To assess the significance of the outliers, we compare the values of all 
the invariants $X_{ij}$ for multipoles 
$\ell=2-20$ to those from simulations: for each value we find the number of simulations 
that find a lower value. We highlight as anomalous any result where either $>$99\% or 
$<$1\% of the simulations found lower values, and we record just these results in 
Table~\ref{anom}.

\begin{table}
\centering

\begin{tabular}{lccr}
\hline
$\ell$ & i & j & \%  \\
\hline
{\bf coadd} & & & \\
\hline
11    &       1      &     5 & 0.8    \\   
12    &       1      &     6 & 1.0    \\  
13    &       1      &     3 & 0.8    \\ 
13    &       2      &     8 & 0.6    \\ 
13    &       1      &     9 &  99.1  \\  
14    &       2      &    14 &  99.7  \\  
15    &       2      &    12 & 0.4    \\  
16    &       1      &    11 &  99.5  \\  
\hline
\hline
{\bf ILC} & & & \\
\hline
 6   &        1     &      2 & 0.9  \\   
11   &        1     &      9 &  99.6 \\   
13   &        1     &      6 &  99.7 \\   
13   &        2     &     10 &  99.5 \\   
16   &        2     &     10 &  99.0 \\   
18   &        1     &      4  & 99.5 \\   
18   &        1     &      6 & 0.2   \\   
18   &        2     &      8 &  99.4 \\   
19   &        1     &      5 &  99.5 \\     
\hline 
\hline
{\bf TOHcl} & & &  \\
\hline
 3      &     2     &      3 & 0.8   \\   
 8      &     2     &      4 &  99.5 \\   
13      &     1     &      6 & 0.4   \\   
14      &     1     &      9 & 0.6   \\   
16      &     2     &     13 &  99.0 \\   
19      &     2     &     10 &  99.4 \\   
\hline
\hline
{\bf TOHw} & & &  \\
\hline
 3    &       2      &     3 & 0.8    \\      
 8    &       2      &     5 &  99.1  \\  
 9    &       1      &     6 & 0.1    \\  
13    &       1      &     5 & 0.2    \\ 
17    &       1      &    14 & 0.5    \\   
18    &       1      &    11 &  1.0   \\   
19    &       2      &     9 &  99.8  \\     
19    &       1      &    18 &  99.0  \\   
20    &       1      &    10 &  1.0   \\   
\hline
\end{tabular}
\caption{Anomalous multipole invariant values, at the 98\% significance level. The percentage of simulations 
that find lower values.}
\label{anom}
\end{table}

Before we take any of these results too seriously, we must note that a certain of 
number of anomalous values 
are \emph{expected}. That is the same with any statistic. For each multipole 
there are $2\ell-3$ invariants that 
we measure. Therefore, when we look at the multipoles $\ell=2-20$ for each map 
we observe a total of 361 invariants. We have highlighted the values outside the 
98\% confidence interval, and 
so we would expect 2\% of our observations to be highlighted: $\sim$7 results for 
every map. We see in the tables that the {\bf coadd} map finds 8 ``anomalies'', 
the {\bf ILC} finds 9,{TOHcl} finds 6, and {\bf TOHw} finds 9. We also expect to find these 
$\sim$7 values in the higher multipoles, as these return more invariants. 

Bearing this in mind, we highlight as particularly interesting the ``anomalies'' 
found in the low multipole 
$\ell=3$, for maps {\bf TOHcl} and {\bf TOHw}. This result shows a particularly low dot 
product between the 
second and third vectors, and therefore an unusually large angle.

Also of significance are the multipoles where 3 ``anomalous'' values are found: 
$\ell=13$ for the {\bf coadd} map, and $\ell=18$ for the {\bf ILC} map.

However, we conclude that through analysis of the Multipole Vectors there is no 
evidence for inherent non-Gaussianity or m-preference in the multipoles $\ell=2-20$.


\section{Discussion}\label{end}
With improving CMB observations, the job of analysing the data also requires improved 
methods. It is unsatisfactory to investigate numerous statistics that bear no intuitive 
investigative properties. In particular, the issues of SI and Gaussianity should 
be probed separately.

In this paper we have applied a method proposed in~\cite{us2} to the WMAP data. The method uses all 
the information in a map, and separates out the issues of SI and Gaussianity by defining 
an orthonormal frame (Multipole Frame) for each multipole, and a set of invariants (Multipole 
Invariants). In theory, the robustness of this method makes it ideal for a thorough investigation 
of anomalies in the CMB.

We analysed the Multipole Vectors of four different CMB maps, all derived from the WMAP data. We 
looked at the multipoles $\ell=2-20$, and for each multipole we found a Multipole Frame and the 
Multipole Invariants. 
To investigate the Statistical Isotropy of the data we analysed the orientation of 
the Multipole Frames over 2 separate multipole 
ranges, $\ell=2-10$, and $\ell=2-20$. We found no evidence that the Euler angles of these 
frames significantly favoured any particular values. Therefore we found no evidence for 
a departure from SI.
To investigate the Non-Gaussianity of the data we analysed all the 361 Multipole Invariants for 
multipoles $\ell=2-20$ and compared their values to those from simulations. We recorded the 
extreme values, and saw that each map found approximately the expected amount. Therefore we found 
no evidence for Non-Gaussianity.

In light of previously detected anomalies~\citep{coles,copi1,us3,oliv,virgo,copi2,teg}, 
the failure of our method to return any 
significant results raises a couple of interesting points. 
The effectiveness of our method comes into question, and in practice we have seen that our 
method is victim to discontinuous noise. The method pivots around the selection of Anchor 
Vectors. Small fluctuations in the positions of the Multipole Vectors alters their 
ordering and therefore has a large effect on our method. The Anchor vectors can then differ, 
and this induces a discontinuous noise in the Euler angles of the Multipole Frames and in the 
Multipole Invariants. Also, different ways of ordering the vectors will make the method 
effective at detecting different features.

Another significant issue raised is that of priors. In this paper we limit ourselves to the 
multipoles $\ell=2-20$, but we do not focus any further or assume priors about any range. 
Therefore we may overlook interesting features in a few multipoles because their anomalous behaviour 
will be diluted by the well behaved values from other multipoles. In only 
focusing on interesting multipoles one has assumed a prior. In this paper we have applied 
our method to look for more overall features.

We conclude that the Multipole Vector method with the anchor vectors is technically 
ideal, but in practice is very limited by noise. In its application we have 
found no evidence for anisotropy or non-gaussianity. However, we feel that in light of 
other reported results this is because our method overlooks subtle features in the 
data. What we gain in thoroughness, we loose in sensitivity. We find no evidence for 
inherent anisotropy or Non-Gaussianity.

\section*{Acknowledgements}
We thank Jeff Weeks, Max Tegmark, and Glenn Starkmann for helpful comments. We also thank 
Joao Medeiros and Andrew Jaffe for interesting discussions. We are grateful 
for the use of the Multipole Vector decomposition computer codes
\citep{copi1}\footnote{Available at www.phys.cwru.edu/projects/mpvectors/}, and 
we used the HEALPix 
package (\cite{healp}\footnote{Available at http://www.eso.org/science/healpix}). 
Calculations were performed on COSMOS, the UK cosmology 
supercomputer facility. KRL is funded by PPARC.

\label{lastpage}


\begin{thebibliography}{99}
\bibitem[\protect\citeauthoryear{Bennett \& al}{2003}]{bennetta}
Bennett C., et al., 2003, Astrophys J Suppl. 148 1.
\bibitem[\protect\citeauthoryear{Bennett \& al}{2003}]{bennettb}
Bennett C., et al., 2003, Astrophys J Suppl. 148 97.
\bibitem[\protect\citeauthoryear{Berera \& al}{2003}]{arj}
Berera A., Buniy R., Kephart T., hep-th/0311223.
\bibitem[\protect\citeauthoryear{Coles \& al}{2003}]{coles}
Coles P., et al., 2003, astro-ph/0310252
\bibitem[\protect\citeauthoryear{Copi, Huterer, Starkman}{2003}]{copi1}
Copi C.J., Huterer D., Starkman G.D., 2003, astro-ph/0310511
\bibitem[\protect\citeauthoryear{Cornish \& al}{2004}]{sperg}
Cornish N., et al., Phys. Rev. Lett. 92, 201302, 2004.
\bibitem[\protect\citeauthoryear{Cruz \& al}{2004}]{cruz}
Cruz M., et al., 2004, astro-ph/0405341
\bibitem[\protect\citeauthoryear{Dennis}{2004a}]{dennis1}
Dennis M.R., 2004a, math-ph/0408046
\bibitem[\protect\citeauthoryear{Dennis}{2004b}]{dennis2}
Dennis M.R., 2004b, math-ph/0410004
\bibitem[\protect\citeauthoryear{Donoghue, Donoghue}{2004}]{dondon}
Donoghue E.P., Donoghue J.F., 2004, astro-ph/0411237
\bibitem[\protect\citeauthoryear{Eriksen \& al}{2003}]{erik1}
Eriksen H.K., et al, 2y003, astro-ph/0307507
\bibitem[\protect\citeauthoryear{Eriksen \& al}{2004a}]{erik2}
Eriksen H.K., et al, 2004a, astro-ph/0401276
\bibitem[\protect\citeauthoryear{Eriksen \& al}{2004b}]{erik3}
Eriksen H.K., et al, 2004b, astro-ph/0407271
\bibitem[\protect\citeauthoryear{Hajian, Souradeep}{2005}]{BiPS}
Hajian A., Souradeep T., 2005, astro-ph/0501001
\bibitem[\protect\citeauthoryear{G\'orski, Hivon, Wandelt}{1999}]{healp}
G\'orski K.M., Hivon E., Wandelt B., 1999, astro-ph/9905275
\bibitem[\protect\citeauthoryear{Hansen \& al}{2004a}]{hansen1}
Hansen F.K. et al., 2004a, astro-ph/0402396
\bibitem[\protect\citeauthoryear{Hansen, Banday, Gorski}{2004}]{hansen2}
Hansen F.K., Banday A.J., Gorski K.M., 2004, astro-ph/0404206
\bibitem[\protect\citeauthoryear{Hu}{2001}]{hu}
Hu W., 2001, astro-ph/0105117
\bibitem[\protect\citeauthoryear{Katz, Weeks}{2004}]{katz}
Katz G., Weeks J., 2004, astro-ph/0405631
\bibitem[\protect\citeauthoryear{Lachi\`{e}ze-Rey}{2004}]{lrey}
Lachi\`{e}ze-Rey M., 2004, astro-ph/0409081
\bibitem[\protect\citeauthoryear{Land, Magueijo}{2004a}]{us1}
Land K.R., Magueijo J., 2004a, astro-ph/0405519
\bibitem[\protect\citeauthoryear{Land, Magueijo}{2004b}]{us2}
Land K.R., Magueijo J., 2004b, astro-ph/0407081
\bibitem[\protect\citeauthoryear{Land, Magueijo}{2005}]{us3}
Land K.R., Magueijo J., 2005, astro-ph/0502237
\bibitem[\protect\citeauthoryear{Larson, Wandelt}{2004}]{larson}
Larson D.L., Wandelt B.D., 2004, astro-ph/0404037
\bibitem[\protect\citeauthoryear{Magueijo}{1994}]{conf}
Magueijo J.C.R., 1994, astro-ph/9412096
\bibitem[\protect\citeauthoryear{Maxwell}{1891}]{maxwell}
Maxwell J.C., 1891, A Treatise on Electricity and Magnetism, Clarendon Press
\bibitem[\protect\citeauthoryear{Moffat}{2005}]{moffatinh}
Moffat J., astro-ph/0502110.
\bibitem[\protect\citeauthoryear{Oliveira-Costa \& al}{2003}]{oliv}
Oliveira-Costa A. et al., 2003, astro-ph/0307282 
\bibitem[\protect\citeauthoryear{Park}{2003}]{park}
Park C., 2003, astro-ph/0307469 
\bibitem[\protect\citeauthoryear{Ralston, Jain}{2004}]{virgo}
Ralston J.P., Jain P., 2004, astro-ph/0311430
\bibitem[\protect\citeauthoryear{Roukema \& al}{2004}]{dodec}
Roukema B.F., et al., astro-ph/0402608.
\bibitem[\protect\citeauthoryear{Schwarz \& al}{2003}]{copi2}
Schwarz D.J. et al., 2003, astro-ph/0403353
\bibitem[\protect\citeauthoryear{Tegmark, Oliveira-Costa, Hamilton}{2003}]{teg}
Tegmark M., Oliveira-Costa A., Hamilton J.S., 2003, Phys.Rev.D68(2003)123523
\bibitem[\protect\citeauthoryear{Vielva \& al}{2003}]{vielva}
Vielva P., et al., 2003, astro-ph/0310273
\bibitem[\protect\citeauthoryear{Weeks}{2004}]{weeks}
Weeks J., 2004, astro-ph/0412231
\bibitem[\protect\citeauthoryear{Weeks \& al}{2004}]{riaz}
Weeks J., et al., Mon. Not. Roy. Astron. Soc. 352, 258, 2004.
\end{thebibliography}
\end{document}